\pdfoutput=1 
\documentclass[aps,pra,twocolumn,groupedaddress,amsmath]{revtex4-1}

\usepackage{graphicx}
\usepackage{amssymb}
\usepackage{bm}
\usepackage{amsmath,amsfonts,latexsym}

\usepackage{braket} 

\usepackage{color}

\usepackage{comment}

\usepackage{bm} 

\newcommand*{\br}{\mathbf{r}}

\newcommand*{\aop}{A^{\vphantom{\dagger}}}
\newcommand*{\adop}{A^\dagger}

\begin{document}

\title{Spin 1 microcondensate in a magnetic field: semiclassics and exact solution}

\author{Austen Lamacraft} 
\affiliation{Department of Physics, University of Virginia,
Charlottesville, VA 22904-4714 USA}
\date{\today}
\email{austen@virginia.edu}
 
\date{\today}

\begin{abstract}

We study a spin 1 Bose condensate small enough to be treated as a single magnetic `domain': a system that we term a \emph{microcondensate}. Because all particles occupy a single spatial mode, this quantum many body system has a well defined classical limit consisting of three degrees of freedom, corresponding to the three macroscopically occupied spin states. We study both the classical limit and its quantization, finding an integrable system in both cases. Depending on the sign of the ratio of the spin interaction energy and the quadratic Zeeman energy, the classical limit displays either a separartrix in phase space, or \emph{Hamiltonian monodromy} corresponding to non-trivial phase space topology. We discuss the quantum signatures of these classical phenomena using semiclassical quantization as well as an exact solution using the Bethe ansatz.

\end{abstract}

\maketitle

\section{Introduction} 
\label{sec:introduction}


A spin 1 Bose condensate can be regarded as an unusual kind of magnet. As with a magnet, we can consider a small system that constitutes a single magnetic `domain'. The study of the dynamics and excitation spectrum of a spin 1 condensate in this regime -- which we term a \emph{microcondensate} -- is the subject of this paper. 


A number of recent experiments have demonstrated the relevance of this `single mode' approximation in experiments using condensates of $^{87}$Rb or $^{23}$Na atoms \cite{Chang:2005,black:2007,liu:2009}. This regime occurs when the size of the condensate is less than the spin healing length over which the spin state of the particles can change significantly. This allows us to consider the spin dynamics in a single spatial mode. In terms of the operators $\adop_{m}$, $\aop_{m}$ $m=-1,0,1$ that create and destroy particles in the three spin states of that mode, the Hamiltonian for a system of $N$ particles is
%
%
%
\begin{equation}\label{SMA}
H_{\text{SMA}}=\frac{e_{0}}{2N}:\hat{N}^{2}:+\frac{e_{2}}{2N}: \hat{\mathbf{S}}\cdot\hat{\mathbf{S}}:+H_{\text{Z}}.
\end{equation}
Here the colons denote normal ordering and $\hat{N}$ and $\hat{\mathbf{S}}$ are respectively the operators of total number and spin 
\begin{equation}
	\label{SpinorCondensateMonodromy_numberspin}
	\hat{N}=\sum_{m=-1}^{1}\adop_{m}\aop_{m}\qquad \hat{\mathbf{S}}=\sum_{m,m'}\adop_{m}\bm{\mathsf{S}}_{mm'}\aop_{m'},
\end{equation}
where $\bm{\mathsf{S}}_{mn}$ are the spin 1 matrices. The energies $e_{0}$ and $e_{2}$ parametrize the strengths of the density-density and spin-spin interactions, the most general situation for spin 1 \footnote{We have $e_{0,2}=c_{0,2}N\int d\br\, |\varphi(\br)|^{4}$ in terms of the microscopic interaction parameters $c_{0,2}$ and $\varphi(\br)$ the condensate wavefunction~\cite{ho1998,ohmi1998}}. $H_{\text{Z}}$ describes the Zeeman energy, and includes both linear and quadratic contributions
\begin{equation}\label{zeeman}
H_{\text{Z}}=\sum_{m=-1}^{1}\adop_{m}\left[p m+q m^{2}\right]\aop_{m}
\end{equation}
The investigation of the Hamiltonian Eq.~\eqref{SMA} is the main goal of this paper. It is integrable both at the classical and quantum level. We shall show that this rather simple system displays a variety of  interesting properties that merit further experimental study. 

When the number $N$ is large we expect on general grounds that the quantities $\aop_{m}$, $\adop_{m}$ can be treated as classical amplitudes $A^{\vphantom{*}}_{m}$, $A^{*}_{m}$, with a classical Hamiltonian given by Eq.~\eqref{SMA}, and Poisson brackets that reflect the more familiar quantum commutators
\begin{equation}
	\label{SpinorCondensateMonodromy_PBs}
	\left\{A^{*}_{m},A^{\vphantom{*}}_{m'}\right\}_{\text{PB}}=\frac{i}{\hbar}\delta_{mm'}.
\end{equation}
The resulting equations of motion are nothing but the `zero-dimensional' Gross--Pitaevskii equation. This mechanical system has three degrees of freedom (so that the phase space is six dimensional, corresponding to the three complex quantities $A^{\vphantom{*}}_{m}$, $m=-1,0,+1$), and three commuting conserved quantities: $\hat{N}$, $\hat{S}_{z}$, and the energy $H_{\text{SMA}}$ itself (note that $\hat{\mathbf{S}}^{2}$ is not conserved because of the quadratic Zeeman effect). As a result, the system is integrable in the sense of Liouville, a notion that will be reviewed in Section~\ref{sub:general_consequences_of_integrability}. 

We will see that the character of the classical motion depends essentially on the sign of $\tilde q\equiv q/e_{2}$. Since $q$ is typically positive (though an effective negative $q$ can be induced by microwave dressing \cite{gerbier:2006,leslie:2009}), this means that $\tilde q>0$ is most naturally realized for antiferromagnetic interactions e.g. in $^{23}\text{Na}$ and $\tilde q<0$ in the ferromagnetic case e.g. $^{87}\text{Rb}$. For a graphical illustration of the difference between these two cases, the reader should compare the spectra shown in Figs.~\ref{fig:qgreater_bndy} and \ref{fig:qless_boundary} (note that the spectra shown are all plotted with $e_{2}=1$, so the spectrum should be inverted in the ferromagnetic case). The way in which the classical motion influences the quantum spectrum is one of the main themes of this paper.



The structure of the remainder of this paper is as follows. In Section~\ref{sec:formulation_in_terms_of_hyperbolic_spins} we recast the Hamiltonian Eq.~\eqref{SMA} in terms of hyperbolic (or $SU(1,1)$) spins, which provides a convenient framework for the analysis of both the classical and quantum systems. Section~\ref{sec:the_classical_limit_reduction_and_monodromy} presents a detailed analysis of the classical problem and its semiclassical quantization. After describing the features of the classical dynamics in qualitative terms and explaining the relation to the mean-field phase diagram, we introduce some ideas from the theory of classical integrable systems, notably action-angle coordinates, that are then applied to the system of interest. It is in this section that we meet the phenomenon of \emph{Hamiltonian monodromy}, a topological obstruction to the global existence of action-angle coordinates (Section~\ref{sub:rotation_number_and_monodromy}) \cite{Duistermaat:1980,cushman:1997,cushman:2000}. This material may be unfamiliar to many readers, so we have tried to be pedagogical in our presentation (other introductions suitable for physicists may be found in the appendices to Ref.~\cite{efstathiou2004}, see also Ref.~\cite{efstathiou2010}). Finally Section~\ref{sec:solution_of_the_quantum_hamiltonian} is devoted to a very different approach, in which the quantum Hamiltonian is solved directly using the Bethe ansatz. The Bethe ansatz equations were given in Ref.~\cite{Bogoliubov:2006}, based on mapping to a type of Gaudin model written in terms of the hyperbolic spins. Here we re-derive these equations by a different method and describe their solution, with a particular emphasis on the properties already uncovered in Section~\ref{sec:the_classical_limit_reduction_and_monodromy}.  

We close this introductory section with a couple of comments on some related work. The low energy physics of spin 1 condensates with antiferromagnetic interactions (also called `polar' condensates) has often been studied in terms of an effective `rotor' description, akin to the low energy description of the N\'eel state of an antiferromagnet \cite{zhou:2001}. Refs.~\cite{barnett:2010a,barnett:2010} recently extended this description to the full spectrum in the single mode approximation. The rotor formulation is quite different from the approach pursued in this work, however. Finally, a model identical to Eq.~\eqref{SMA} was studied numerically, and the existence of monodromy pointed out, in Ref.~\cite{Perez-Bernal:2008} in a very different context. Our goal here is to provide an \emph{analytic} description.


\section{Formulation in terms of hyperbolic spins} 
\label{sec:formulation_in_terms_of_hyperbolic_spins}

Let us write  Eq.~\eqref{SMA} explicitly in terms of the boson field operators. For clarity we drop the terms proportional to $e_{0}$ (density-density interaction) and $p$ (linear Zeeman effect) as these couple to conserved quantities, giving the reduced Hamiltonian
\begin{widetext}
	\begin{align}
		\label{SpinorCondensateMonodromy_SimpleHam}
		 H_{\text{red}}&
 =\frac{e_{2}}{2N}\left[S_{z}^{2}+2\left(\adop_{1}\adop_{-1}(\aop_{0})^{2}+(\adop_{0})^{2}\aop_{1}\aop_{-1}\right)+2\left(\adop_{0}\aop_{0}-\frac{1}{2}\right)\left(\adop_{1}\aop_{1}+\adop_{-1}\aop_{-1}\right)\right]
		+q\left[\adop_{1}\aop_{-1}+\adop_{-1}\aop_{-1}\right].
	\end{align}	
\end{widetext}
In this form the Hamiltonian is still somewhat indigestible. A considerable simplification was achieved in Ref.~\cite{Bogoliubov:2006} by introducing the variables
\begin{equation}\label{hyp_spins}
\begin{split}
	K_0&=\frac{1}{2}\left[\adop_1\aop_1+\adop_{-1}\aop_{-1}+1\right]\\
	K_+&=\adop_1\adop_{-1},\qquad K_-=\aop_1\aop_{-1}, \\ 
	B_{0}&=\frac{1}{2}\adop_{0}\aop_{0}+\frac{1}{4}, \\
	B_+&=- \frac{1}{2}\left(\adop_0\adop_{0}\right),\, B_-=-\frac{1}{2}\left(\aop_0\aop_{0}\right).	
\end{split}
\end{equation}
We make two important comments about this choice of variables. Firstly, they are invariant under the rotations about the $z$-axis generated by the conserved quantity $S_{z}$ (note that from now on we will we drop the tildes on $N$ and $S_{z}$, trusting that no confusion between the quantum operators and their eigenvalues will result), and are therefore suited to exploiting this symmetry of the system. Secondly, they constitute two representations of the non-compact group $SU(1,1)$, obeying the relations
\begin{subequations}
\label{SpinorMonod_SU(1,1)quant}
\begin{align}
	\left[K_{0},K_{\pm}\right]&=\pm K_{\pm}\\
	\left[K_{+},K_{-}\right]&=-2K_{0}	\label{su(1,1)2} 
\end{align}
\end{subequations}
and likewise for the $\left\{B_{0},B_{+},B_{-}\right\}$ variables. The difference from the more familiar $SU(2)$ algebra lies in the minus sign in  Eq.~\eqref{su(1,1)2}, leading to the quadratic Casimir operators
\begin{equation}
	\label{SpinorMonod_casimirs}
\begin{split}
	C_{K}&=K_{0}^{2}-\frac{1}{2}\left(K_{-}K_{+}+K_{+}K_{-}\right)=\frac{1}{4}\left(S_{z}^{2}-1\right)\\
	C_{B}&=B_{0}^{2}-\frac{1}{2}\left(B_{-}B_{+}+B_{+}B_{-}\right)=-\frac{3}{16}.
\end{split}
\end{equation}
In the first case the representation is fixed by specifying the value of $S_{z}=\adop_{1}\aop_{1}-\adop_{-1}\aop_{-1}$. In terms of the occupation number basis $\ket{N_{1}}_{1}\ket{N_{-1}}_{-1}$ of the $m=\pm 1$ states the highest weight state is
\begin{equation}
	\label{SpinorMonod_highest}
\begin{split}
	\ket{0,\nu_{K}}_{K}\equiv
	\begin{cases}
		\ket{{S}_{z}}_{1}\ket{0}_{-1},&\text{ if ${S}_{z}\geq 0$}\\
		\ket{0}_{1}\ket{-{S}_{z}}_{-1},&\text{ if ${S}_{z}\leq 0$}
	\end{cases}
\end{split}
\end{equation}
where the Bargmann index $\nu_{K}\equiv \frac{1}{2}\left(|S_{z}|+1\right)$ is the eigenvalue of $K_{0}$ for the highest weight state. Note that $C_{K}=\nu_{K}\left(\nu_{K}-1\right)$. Repeated application of $K_{+}$ generates the states
\begin{equation}
	\label{SpinorMonod_Kstates}
\begin{split}
	K_{+}\ket{n,\nu_{K}}_{K}=\sqrt{(2\nu_{K}+n)(n+1)}\ket{n+1,\nu_{K}}_{K}\\
	\ket{n,\nu_{K}}_{K}\equiv
	\begin{cases}
		\ket{S_{z}+n}_{1}\ket{n}_{-1},&\text{ if $S_{z}\geq 0$}\\
		\ket{n}_{1}\ket{-S_{z}+n}_{-1},&\text{ if $S_{z}\leq 0$}
	\end{cases}
\end{split}
\end{equation}
with $K_{0}$ eigenvalue
\[
	K_{0}\ket{n,\nu_{K}}_{K}=(n+\nu_{K})\ket{n,\nu_{K}}.
\]
For the $\left\{B_{0},B_{+},B_{-}\right\}$ variables (`one mode representation') things are slightly different. There are just two representations, with highest weight state $\ket{0}_{0}$ or $\ket{1}_{0}$ i.e. either no particle or one particle in the $m=0$ state. These have $B_{0}=\frac{1}{4},\frac{3}{4}$, so
\begin{equation}
	\label{SpinorMonod_highestB}
\begin{split}
	\ket{0,\nu_{B}=1/4}_{B}\equiv \ket{0}_{0}\\
		\ket{0,\nu_{B}=3/4}_{B}\equiv \ket{1}_{0}
\end{split}	
\end{equation}
both giving $C_{B}=\nu_{B}\left(\nu_{B}-1\right)=-\frac{3}{16}$. This index is determined by the parity of $N-S_{z}$
\begin{equation}
	\label{SpinorMonod_parity}
\begin{split}
	\nu_{B}=
	\begin{cases}
		\frac{1}{4}, &\text{for $N-S_{z}$ even}\\
		\frac{3}{4}, &\text{for $N-S_{z}$ odd}
	\end{cases}
\end{split}
\end{equation}
Repeated application of $B_{+}$ generates the states
\begin{equation}
	\label{SpinorMonod_Bstates}
\begin{split}
	B_{+}\ket{n,\nu_{B}}_{B}=\sqrt{(2\nu_{B}+n)(n+1)}\ket{n+1,\nu_{B}}_{B}\\
	\ket{n,\nu_{B}}_{B}\equiv
	\begin{cases}
		(-1)^{n}\ket{2n}_{0},&\text{ if $\nu_{B}=\frac{1}{4}$}\\
		(-1)^{n}\ket{2n+1}_{0},&\text{ if $\nu_{B}=\frac{3}{4}$}.
	\end{cases}
\end{split}
\end{equation}
In the classical limit we can think of the Casimirs of Eq.~\eqref{SpinorMonod_casimirs} as specifying a hyperboloid in the space of values $\left(K_{x},K_{y},K_{z}\right)$ with \footnote{More precisely, we are concerned with the upper sheet of a two-sheeted hyperboloid}
\begin{equation}
	\label{SpinorMonod_HypSpace}
\begin{split}
	 K_{x}&=\frac{1}{2}\left(K_{+}+K_{-}\right)\\
	K_{y}&=\frac{1}{2i}\left(K_{+}-K_{-}\right)\\
	 K_{z}&=K_{0}	
\end{split}
\end{equation}
and similarly for the $B$ representation. Thus we will refer to these variables as `hyperbolic spins'. In this language the Hamiltonian Eq.~\eqref{SpinorCondensateMonodromy_SimpleHam} takes the form
%
%
	\begin{equation}
		\label{SpinorCondensateMonodromy_HypSpinHam}
		 H_{\text{red}}
 =\frac{e_{2}}{2N}\left[S_{z}^{2}+8B_{0}K_{0}-4B_{-}K_{+}-4B_{+}K_{-}\right]
		+2q K_{0}.
	\end{equation}	
%
We have dropped a term $-e_{2}\left(1+\frac{1}{2N}\right)-q$, constant by number conservation. In Eq.~\eqref{SpinorCondensateMonodromy_HypSpinHam} the conservation of particle number $N=2(K_{0}+B_{0})-\frac{3}{2}$ is also manifest as a symmetry under rotations about the $z$-axis in the space of the hyperbolic spins (see Eq.~\eqref{SpinorMonod_HypSpace}). We readily see that in terms of the original $\adop_{m}$ degrees of freedom, this is just the overall phase conjugate to the total particle number.

We emphasize that Eq.~\eqref{SpinorCondensateMonodromy_HypSpinHam} is identical to the original Hamiltonian Eq.~\eqref{SpinorCondensateMonodromy_SimpleHam}. The symmetry under rotations about the $z$-axis (equivalently, conservation of $S_{z}$) allows us to write the original problem with three degrees of freedom as a problem with only two degrees of freedom. The price we pay is that the relevant variables are the less familiar hyperbolic spins.

This formulation of the problem turns out to be a very convenient starting point for the analysis of both the classical and quantum dynamics of the spin 1 condensate. 



\section{The (semi-)classical limit: reduction and monodromy} 
\label{sec:the_classical_limit_reduction_and_monodromy}

\subsection{Qualitative features of reduced dynamics} 
\label{sub:qualitative_features_of_dynamics}

In this section we analyze the classical mechanics of the system described by Eq.~\eqref{SpinorCondensateMonodromy_HypSpinHam}. On general grounds we expect this to be a good description when the number of particles $N$ is large, in which case the operators $\aop_{m}$, $\adop_{m}$ can be treated as classical amplitudes  $A^{\vphantom{*}}_{m}$, $A^{*}_{m}$ of `order $\sqrt{N}$'. Thus the $B$ and $K$ variables are $O(N)$, as is $S_{z}$, so in this limit we take
\begin{equation}
	\label{SpinorMonod_ClassCasimirs}
\begin{split}
	C_{K}&\sim \frac{S_{z}^{2}}{4} \\
	C_{B}&\sim 0. \\
	N &\sim 2\left(B_{0}+K_{0}\right)
\end{split}
\end{equation}
and the $O(1)$ terms in Eq.~\eqref{hyp_spins} can be dropped. Thus the $B$ degrees of freedom are restricted to a cone $B_{0}=|B_{+}|$. Note that $K_{0},B_{0}\leq \frac{N}{2}$.

We can exploit the symmetry under rotations about the $z$-axis in hyperbolic spin space, corresponding to conservation of particle number, by taking $B_{+}$ to be real and positive. Then we have $B_{+}=B_{0}=N/2-K_{0}$ and we can eliminate the $B$ degrees of freedom from Eq.~\eqref{SpinorCondensateMonodromy_HypSpinHam} to give the classical Hamiltonian
\begin{equation}
	\label{SpinorMonod_HypSpinReduced}
		 H_{\text{class}} =\frac{e_{2}}{2N}\left[S_{z}^{2}+8\left(N/2-K_{0}\right)\left(K_{0}-K_{x}\right)\right]
		+2q K_{0}.
\end{equation}
If the reader finds these manipulations too cavalier, an alternative is to use conservation of $N$ to write the quantum Hamiltonian Eq.~\eqref{SpinorCondensateMonodromy_HypSpinHam} as
\begin{widetext}
	\begin{multline}
		\label{SpinorMonod_QuantumReduced}
		H_{\text{red}} =\frac{e_{2}}{2N}\left[S_{z}^{2}+8\left(\frac{N}{2}-\frac{3}{4}-K_{0}\right)K_{0}-\right.\\ \left.2\sqrt{(N-2K_{0})(N-2K_{0}+1)}K_{+}-2K_{-}\sqrt{(N-2K_{0})(N-2K_{0}+1)}\right]+2q K_{0},
	\end{multline}	
\end{widetext}
which reduces to Eq.~\eqref{SpinorMonod_HypSpinReduced} when the $O(1)$ terms are dropped.

Now of course, the semi-classical limit is the domain of validity of the conceptually simpler framework of the Gross--Pitaevskii equation (see Ref.~\cite{Zhang:2005}, for instance). Why go to all this effort? The answer is that we have exploited the conservation laws of the problem to reduce the dynamics to a single degree of freedom, which allows us to say a good deal about the character of the motion without detailed calculations. We will refer to the hyperboloid of fixed $S_{z}$ with $K_{0}<N/2$ as the \emph{reduced phase space}.

\begin{figure}
	\centering
		\includegraphics[width=\columnwidth]{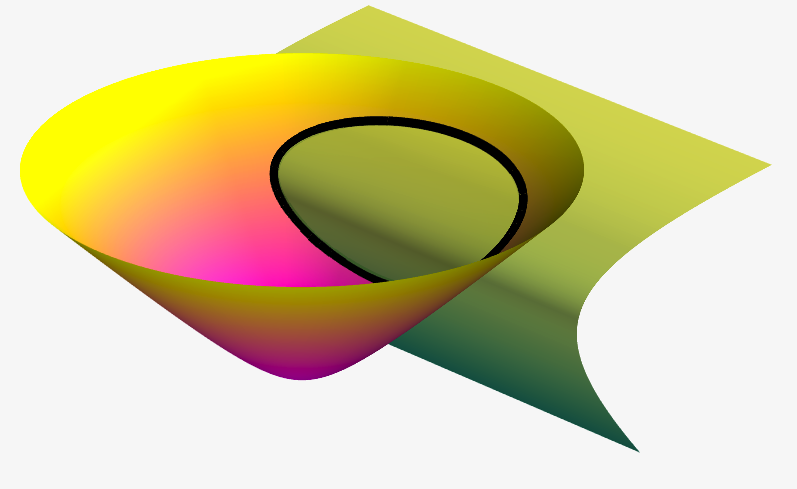}
	\caption{Example of the geometric construction for $\tilde q =0.5$ giving the trajectory (dark line) as the intersection of a hyperboloid, the reduced phase space of fixed $s_{z}=0.2$, and the surface of fixed energy $h=0.7$.}
	\label{fig:geom}
\end{figure}

For the remainder of this section we work with `per particle' quantities in lowercase: thus $k_{x}\equiv K_{x}/N$, $(a_{1},a_{0},a_{-1})$ is the condensate wavefunction normalized to unity, the dimensionless energy per particle is $h\equiv H_{\text{class}}/N e_{2}$, and the dimensionless quadratic Zeeman shift is $\tilde q\equiv q/e_{2}$. Thus
\begin{subequations}
	\label{SpinorMonod_dimless}
\begin{align}
	h &= \frac{1}{2}s_{z}^{2}+2(1-2k_{0})\left(k_{0}-k_{x}\right)+2\tilde q k_{0} \label{h_func}\\
	s_{z} &= \pm 2\sqrt{k_{0}^{2}-|k_{+}|^{2}}
\end{align}
\end{subequations}
Let us begin by discussing the level sets of the Hamiltonian function Eq.~\eqref{h_func}. This is just a quadratic form, which we may write
\begin{widetext}
\begin{equation}
	\label{SpinorMonod_EQuadForm}
	h=\frac{1}{2}s_{z}^{2}+2\begin{pmatrix}
		k_{0}-1/2, k_{x}+\frac{1}{2}(\tilde q-1)
	\end{pmatrix}
	\begin{pmatrix}
		-2 & 1 \cr
		1 & 0
	\end{pmatrix}
	\begin{pmatrix}
		k_{0}-1/2 \\
		k_{x}+\frac{1}{2}(\tilde q-1)
	\end{pmatrix}
+ \tilde q
\end{equation}
\end{widetext}
The surface defined by $h=\text{const.}$ is a hyperbolic sheet whose normal lies in the $(k_{0},k_{x})$ plane. The asymptotes of this surface are the planes $k_{0}=\frac{1}{2}$ and $k_{0}=k_{x}+\frac{\tilde q}{2}$. Theses planes correspond to the vanishing of the quadratic form in Eq.~\eqref{SpinorMonod_EQuadForm} and hence to the energy $h=\frac{1}{2}s_{z}^{2}+\tilde q$. The trajectories of the reduced Hamiltonian are given by the intersection of the hyperboloid of fixed $s_{z}$ with the surface of fixed $h$ (see Fig.~\ref{fig:geom}). 

The character of the trajectories depends critically on the sign of $\tilde q$, and we discuss the two possibilities in turn.

\subsubsection{$\tilde q>0$} 
\label{ssub:qgreater}

\begin{figure}
	\centering
		\includegraphics[width=\columnwidth]{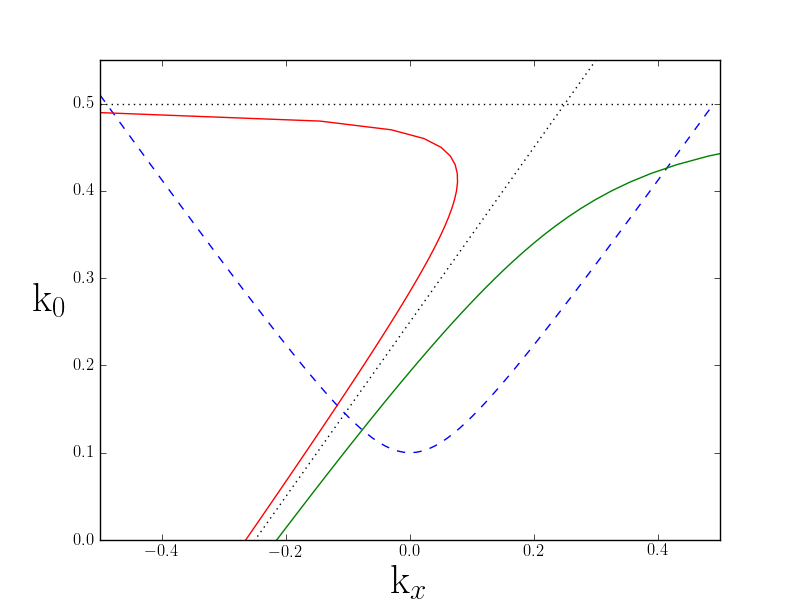}
	\caption{For positive $\tilde q$ (here $\tilde q =0.5$) the asymptotic planes $k_{0}=\frac{1}{2}$, $k_{0}=k_{x}+\frac{\tilde q}{2}$ (black dotted line) of the hyperbolic sheets of constant energy cut the hyperboloid  of fixed $s_{z}=0.2$ (blue dashed line), forming a separatrix between energies greater (red line) or less (green line) than $h=\frac{1}{2}s_{z}^{2}+\tilde q$.}
	\label{fig:qgreater}
\end{figure}

For $0<q<2$ the asymptotic plane $k_{0}=k_{x}+\frac{\tilde q}{2}$ slices through the hyperboloid below $k_{0}=1/2$. Thus, the intersection of the asymptotes with the hyperboloid forms a separatrix at energy $h=\frac{1}{2}s_{z}^{2}+\tilde q$  (see Fig.~\ref{fig:qgreater}). Since $b_{0}=\frac{1}{2}-k_{0}$ it is clear that the separatrix corresponds to a trajectory for the $(b_{x},b_{y},b_{z})$ variables that passes over the tip of the cone $b_{0}=|b_{+}|$.

For a fixed $s_{z}$, only a certain range of values of $h$ are allowed, corresponding to non-zero intersection of the $h=\text{const.}$ surface with the hyperboloid. The endpoints of this range correspond to the stationary points of $h$ on the hyperboloid of fixed $s_{z}$, which are obtained from the equations
\begin{equation}
	\label{SpinorCondensateMonodromy_Highesth}
\begin{split}
	\frac{\partial(h-\lambda s_{z}^{2})}{\partial k_{0}}&=4k_{x}+2-8k_{0}+2\tilde q - 8\lambda k_{0}=0  \\
	\frac{\partial(h-\lambda s_{z}^{2})}{\partial k_{x}}&=2(1-2k_{0})+8 \lambda k_{x}=0,
\end{split}
\end{equation}
where $\lambda$ is a Lagrange multiplier. This pair of linear equations gives $k_{+}$, $k_{0}$ as a function of $\lambda$
\begin{equation}
	\label{SpinorCondensateMonodromy_ksol}
	\begin{pmatrix}
		k_{0}\\
		k_{x}
	\end{pmatrix} = \frac{1}{2(1+2\lambda)^{2}}
	\begin{pmatrix}
		1 + 2 \lambda(1+\tilde q)\\
		1+2\lambda-\tilde q .
	\end{pmatrix}
\end{equation}
These values can be inserted into $h(k_{x},k_{0})$ and $s_{z}(k_{x},k_{0})=2\sqrt{k_{0}^{2}-k_{x}^{2}}$ to yield a parametric curve. 

\begin{figure}
\includegraphics[width=\columnwidth]{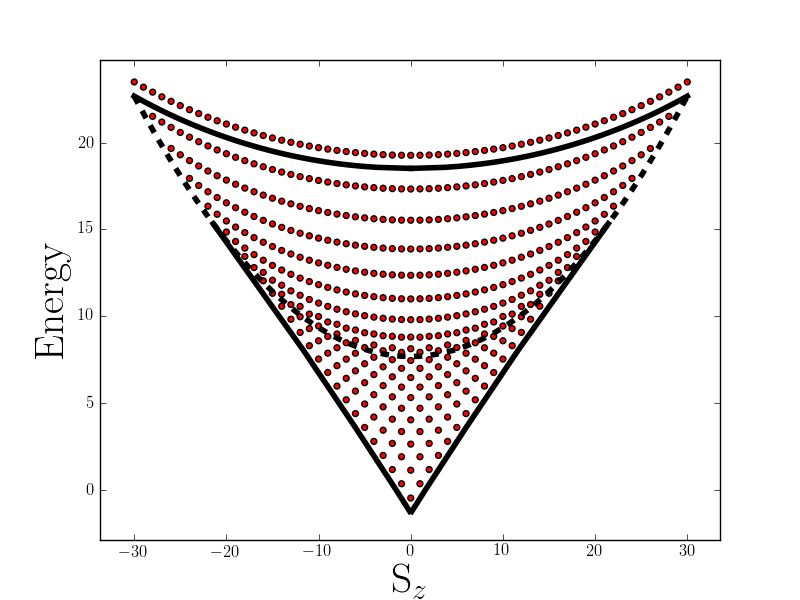}
	\caption{Spectrum of Eq.~\eqref{SpinorCondensateMonodromy_SimpleHam} for $N=30$ particles with $\tilde q=0.3$. The solid lines correspond to the boundaries given by $(s_{z}(k_{x},k_{0}),h(k_{x},k_{0}))$ evaluated on $k_{x,0}(\lambda)$ from Eq.~\eqref{SpinorCondensateMonodromy_ksol} for the ranges in Eq.~\eqref{SpinorMonod_ranges}. The dashed line corresponds to the separatrix $h=\frac{1}{2}s_{z}^{2}+\tilde q$.}
	\label{fig:qgreater_bndy}
\end{figure}

There are two ranges of $\lambda$ that correspond respectively to the upper and lower limiting values of $h$ for each $s_{z}$. 
%
\begin{equation}
	\label{SpinorMonod_ranges}
\begin{split}
	&\text{upper range:}\qquad \frac{\tilde q - 2}{4 + 2\tilde q}<\lambda<\frac{\tilde q -1}{2} \\
	&\text{lower range:}\qquad 0<\lambda<\infty	
\end{split}
\end{equation}
For $\tilde q<1$ this does not include the full lower boundary because for $\frac{\tilde q -1}{2}<\lambda<0$ the solution in Eq.~\eqref{SpinorCondensateMonodromy_ksol} corresponds to values $k_{0}>1/2$. In this case part of the lower boundary is given by the separatrix (see Fig.~\ref{fig:qgreater_bndy}). Once $\tilde q>2$ the \emph{upper} boundary is given by the separatrix.

\subsubsection{$\tilde q<0$} 
\label{ssub:qless}

\begin{figure}
	\includegraphics[width=\columnwidth]{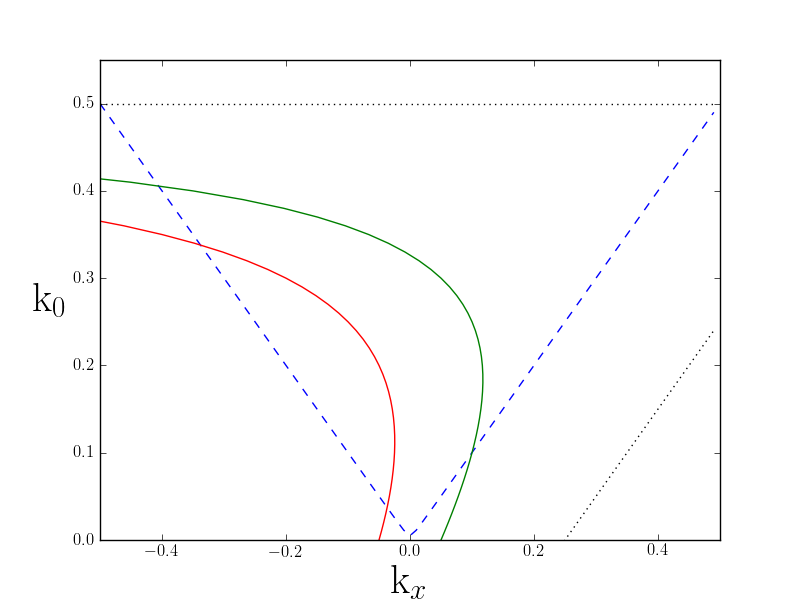}
	\caption{(For negative $\tilde q$ (here $\tilde q =-0.5$) there are two distinct classes of trajectories, that either do (green line $h<0$) or do not (red line $h>0$) encircle the apex of the cone corresponding to $s_{z}=0$. 
	\label{fig:qlesser}}
\end{figure}

In this case the asymptotic plane $k_{0}=k_{x}+\frac{\tilde q}{2}$ never cuts the hyperboloid. The lower energy boundary corresponds to trajectories on the asymptotic plane $k_{0}=\frac{1}{2}$ with $h=\frac{1}{2}s_{z}^{2}+\tilde q$. The upper boundary is found as in the $\tilde q>0$ case.

\begin{figure}
	\centering
		\includegraphics[width=\columnwidth]{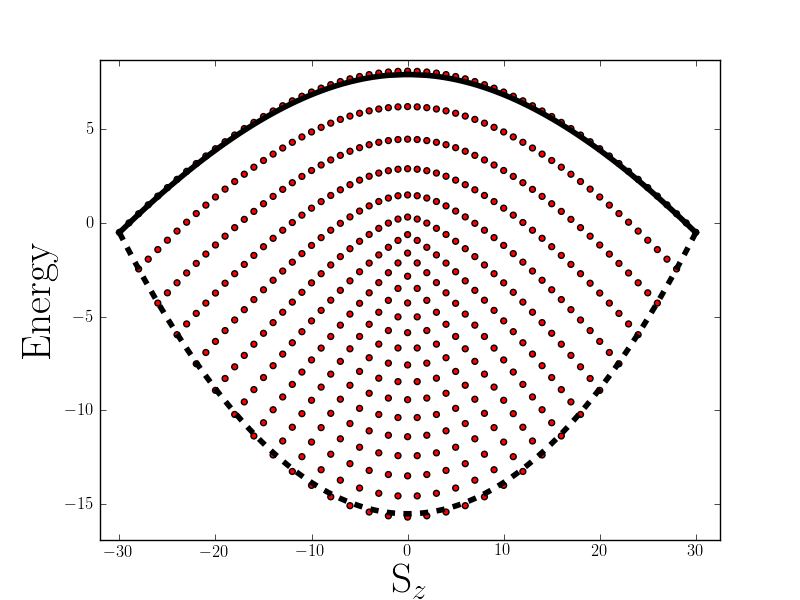}
	\caption{Spectrum of Eq.~\eqref{SpinorCondensateMonodromy_SimpleHam} for $N=30$ particles with $\tilde q=-0.5$. The solid (upper boundary) is found as in the $\tilde q>0$ case, and the lower boundary (dashed) corresponds to trajectories on the asymptotic plane $k_{0}=\frac{1}{2}$ of energy $h=\frac{1}{2}s_{z}^{2}+\tilde q$. The feature at the origin is a signature of the phenomenon of monodromy.}
	\label{fig:qless_boundary}
\end{figure}

An interesting situation does arise, however, for $s_{z}=0$. Here the hyperboloid of fixed $s_{z}$ becomes a cone $k_{0}=|k_{+}|$ and we can distinguish trajectories based upon whether or not they encircle its tip (see Fig.~\ref{fig:qlesser}). From Eq.~\eqref{h_func} we see that the critical value $h=0$ separates these two regimes, with $h<0$ encircling the tip and $h>0$ not. The feature at the origin of the spectrum shown in Fig.~\ref{fig:qless_boundary} is a signature of the phenomenon of monodromy to be discussed shortly, which is related to this distinction.

\begin{figure}
	\centering
		\includegraphics[width=\columnwidth]{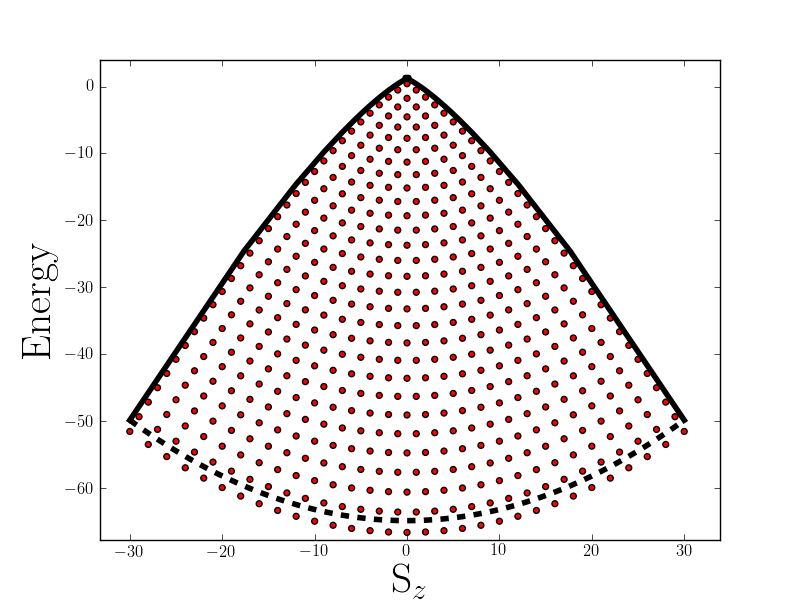}
	\caption{Spectrum of Eq.~\eqref{SpinorCondensateMonodromy_SimpleHam} for $N=30$ particles with $\tilde q=-2.2$. For $\tilde q<-2$ the origin passes through the upper boundary, which becomes cusped.}
	\label{fig:qleast_boundary}
\end{figure}

When $\tilde q<-2$ the origin passes through the upper boundary (note the behavior of the lower limit in Eq.~\eqref{SpinorMonod_ranges}), after which the upper boundary is cusped (see Fig.~\ref{fig:qleast_boundary}). The relation of these changes in the morphology of the spectrum to the mean-field phase diagram will be discussed in the next section. 


\subsubsection{Relation to mean-field phase diagram} 
\label{ssub:relation_to_mean_field_phase_diagram}

The mean-field phase diagram for the ground state in terms of the linear and quadratic Zeeman shifts $p$ and $q$ was given in Ref.~\cite{stenger:1998}. In fact, $p$ is more properly regarded as a Lagrange multiplier used to find the ground state for fixed $S_{z}$. Some of the features discussed in the preceding two sections can be related to the structure of the phase diagram.

\begin{figure}
	\centering
		\includegraphics[width=\columnwidth]{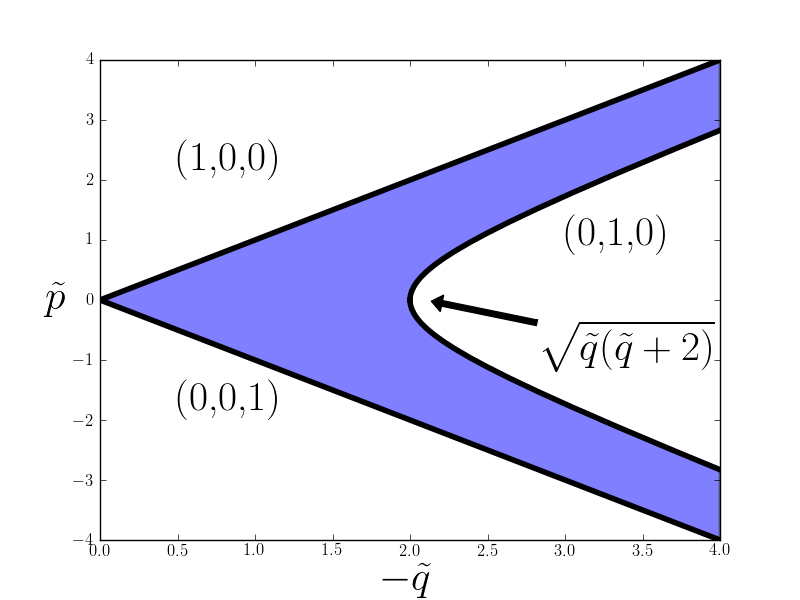}
	\caption{The ground state phase diagram for $e_{2}<0$. The vectors denote the spinor $\left(a_{1},a_{0}, a_{-1}\right)$. The shaded region corresponds to the dark upper boundary in Fig.~\ref{fig:qless_boundary} and Fig.~\ref{fig:qleast_boundary}}
	\label{fig:ferro_phase}
\end{figure}

\begin{figure}
	\centering
		\includegraphics[width=\columnwidth]{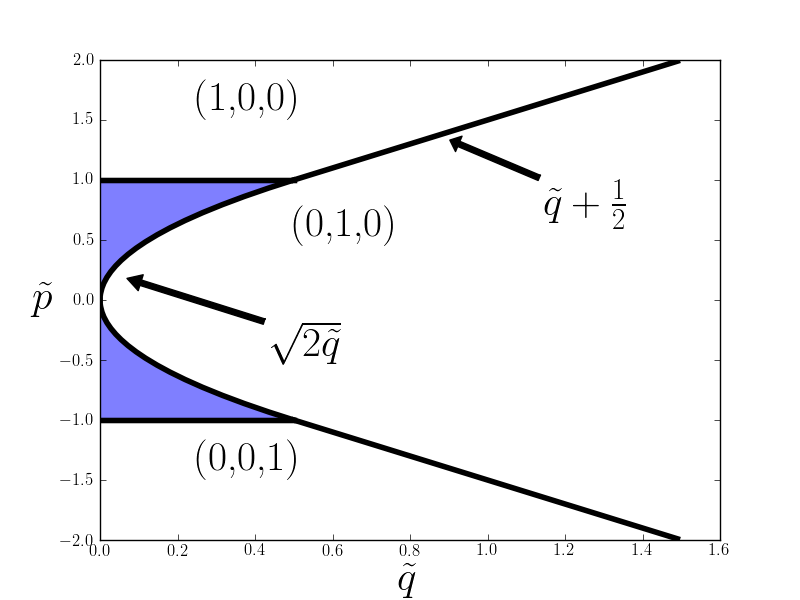}
	\caption{The ground state phase diagram for $e_{2}>0$. The shaded region is where the separatrix (dashed line in in Fig.~\ref{fig:qgreater_bndy}) is the lowest energy.}
	\label{fig:polar_phase}
\end{figure}

Let us begin with the ferromagnetic case, corresponding to $e_{2}<0$. The phase diagram is shown in Fig.~\ref{fig:ferro_phase}. Positive $q$ (the most physically relevant case) corresponds to $\tilde q<0$, and the ground state to the \emph{largest} value of $h$ (since $H_{\text{class}}=h/e_{2}N$). Thus as $p$ increases we move along the dark upper boundary in Fig.~\ref{fig:qless_boundary} and Fig.~\ref{fig:qleast_boundary}, corresponding to the shaded region in Fig.~\ref{fig:ferro_phase}. The difference between $\tilde q >-2$ and $\tilde q<-2$ is that in the latter case a finite $p$ is required before we depart from $s_{z}=0$ because of the cusp in Fig.~\ref{fig:qleast_boundary}. The transition corresponding to the lower boundary of the shaded region in Fig.~\ref{fig:ferro_phase} is the zero-dimensional analog of the transition discussed in Ref.~\cite{lamacraft2007}.

In the antiferromagnetic (or polar) case $e_{2}>0$, we follow the lower boundary in Fig.~\ref{fig:qgreater_bndy}. Because the cusp is always present for $\tilde q >0$ a finite $p$ is always required to depart from $s_{z}=0$. The negative curvature of the black lower boundary means that the ground state jumps straight to a point on the dashed (separatrix) line when $|p|$ exceeds the critical value $\sqrt{2q e_{2}}$. For $\tilde q> \frac{1}{2}$, the system jumps straight to full polarization at $p=q + \frac{e_{2}}{2}$ (see Fig.~\ref{fig:polar_phase}).



\subsection{Classical integrability and its consequences} 
\label{sub:general_consequences_of_integrability}

We began with a system with three degrees of freedom and have reduced the problem to the motion along constant energy contours on a hyperboloid. This reduction used the conservation of $N$ and $S_{z}$, which together with the energy constitute three independent commuting conserved quantities. On general grounds, once we fix the values of these conserved quantities, the system must move on a three dimensional submanifold of the six dimensional phase space. The Liouville--Arnol'd theorem provides more detail, telling us that this submanifold is in fact a three-torus $\mathbb{T}^{3}=S^{1}\times S^{1}\times S^{1}$ \cite{arnold:1989}. The three circles correspond to the overall phase of the spinor $(A_{1},A_{0},A_{-1})$ (conjugate to $N$), the angle of rotation about the $z$-axis (conjugate to $S_{z}$), and the closed trajectories of constant energy on the reduced phase space

The constructive part of the Liouville--Arnol'd provides a distinguished set of coordinates for this torus, known as \emph{action-angle coordinates}. As this idea plays an important role in what follows, we explain it in some detail. Each of the conserved quantites (we denote them by $F_{i}$ $i=1,\ldots N$, for $N$ degrees of freedom) generates a `time' evolution on the torus, by using each in place of the Hamiltonian $H=F_{1}$ in Hamilton's equations. Since these flows commute, the trajectory of a point $x$ on the torus under their combined action can be written $x(t_{1},\ldots, t_{N})$, with the first argument corresponding to the usual time evolution. Now the evolution in each variable is not in general periodic, but rather \emph{\text{quasiperiodic}}, consisting of $N$ incommensurate frequencies. The set of values of $\mathbf{t}=(t_{1},\ldots, t_{N})$ for which $x(t_{1},\ldots, t_{N})=x(0,\ldots,0)$ is a lattice (the \emph{period lattice}) consisting of integer linear combinations of some basis set $\mathbf{e}_{1},\ldots,\mathbf{e}_{N}$, . We can then define the angular variables in terms of the reciprocal lattice vectors $\bm{\epsilon}_{1},\ldots, \bm{\epsilon}_{N}$ satisfying $\bm{\epsilon}_{i}\cdot \mathbf{e}_{j}=2\pi \delta_{ij}$
\begin{equation}
	\label{SpinorMonod_angles}
	\varphi_{i} \equiv \bm{\epsilon}_{i}\cdot \mathbf{t},\qquad i=1,\ldots, N,
	\end{equation}
which increase by $2\pi$ as we advance one unit in each of the lattice directions. These provide a natural parametrization of the torus. Their time evolution due to the Hamiltonian is particularly simple:
\begin{align}
	\label{SpinorMonod_angles_evol}
	\varphi_{i}(t_{1}) &= \omega_{i}t_{1}+\varphi_{i}(0) \nonumber\\
	\omega_{i}&\equiv (\bm{\epsilon}_{i})_{1},\qquad i=1,\ldots, N.
	\end{align}
Note that for a given period lattice, the lattice vectors $\mathbf{e}_{1},\ldots,\mathbf{e}_{N}$ are not unique, resulting in an arbitrariness in the angles that will be important in the following.

The final step is the introduction of the \emph{actions}
\begin{equation}
	\label{SpinorMonod_actions}
	I_{i}\equiv\frac{1}{2\pi} \oint_{\gamma_{i}} \mathbf{p}\cdot d \mathbf{q}_{i},
\end{equation}
where $\mathbf{p}$ and $\mathbf{q}$ are the canonical momentum and position variables, and the integral is taken around the $i^{\text{th}}$ circle of the torus. The definition Eq.~\eqref{SpinorMonod_actions} gives the $I_{i}$ in terms of the conserved quantities $F_{i}$, as specification of the latter fixes the torus. The relation may be inverted to give $F_{i}$, notably the Hamiltonian, in terms of the $I_{i}$.

The actions are conjugate variables to the angles introduced above, so that 
\begin{equation}
	\label{SpinorMonod_Hderiv}
	\omega_{i}=\left(\frac{\partial H}{\partial I_{i}}\right)_{I_{k}\neq I_{i}\text{ fixed}}.
\end{equation}
The evolution generated by each of the conserved quantities corresponds to a matrix of angular `velocities' 
\begin{equation}
	\label{SpinorMonod_AngVelMatrix}
(\bm{\epsilon}_{i})_{j}=\left(\frac{\partial F_{j}}{\partial I_{i}}\right)_{I_{k}\neq I_{i}\text{ fixed}},
\end{equation}
or equivalently the period lattice vectors
\begin{equation}
	\label{SpinorMonod_PeriodLattice}
	(\mathbf{e}_{i})_{j} = 2\pi\left(\frac{\partial I_{i}}{\partial F_{j}} \right)_{F_{k}\neq F_{j}\text{ fixed}}.
\end{equation}
While one can add arbitrary constants to the actions without changing anything, there is more freedom in the choices of angles, where we may redefine 
\begin{equation}
	\label{SpinorMonod_anglegauge}
	\varphi_{i}\to \varphi_{i}+ \Lambda_{i}(I_{1},\ldots, I_{N}).	
\end{equation}
There is a close analogy to the gauge transformations familiar in quantum mechanics. In more mathematical terms, the phase space of an integrable system has the form of a \emph{fiber bundle}, with the base manifold consisting of the space of conserved quantities $F_{i}$, and the fibers being the tori. As we move around the base manifold, we can change the definition of the angular variables on each torus arbitrarily. As in the Aharonov--Bohm effect, interesting things can happen when we move around a circuit containing a singular point, leading to multivaluedness of the angles. This provides one view on the monodromy that we will discuss in Section~\ref{sub:rotation_number_and_monodromy}.

From these generalities we return now to the system of interest. In this case the canonical form $\mathbf{p}\cdot d\mathbf{q}$ becomes $\frac{i}{2} \sum_{m}A^*_{m}dA_{m}-A_{m} dA^{*}_{m}$ (we set $\hbar=1$ for the remainder of this section). Now we choose the following parametrization for the spinor (for $s_{z}>0$)
\begin{equation}
	\label{SpinorMonod_spinorparam}
	\begin{split}
		A_{1}&=\sqrt{S_{z}}\cosh \frac{\theta}{2}e^{-i(\psi/2+\phi+\chi)} \\
		A_{0}&=\sqrt{N_{0}}e^{-i\chi} \\
		A_{-1}&=\sqrt{S_{z}}\sinh \frac{\theta}{2}e^{-i(\psi/2-\phi+\chi)}.		
	\end{split}
\end{equation}
Note that $\chi$ is the overall phase, while $\phi$ describes rotations about the $z$-axis generated by $S_{z}$, and
\begin{equation}
\begin{split}
	K_{+}&=A_{1}^{*}A_{-1}^{*}=\frac{S_{z}}{2}e^{i\psi}\sinh\theta \\
	K_{0}&=\frac{1}{2}\left(|A_{1}|^{2}+|A_{-1}|^{2}\right)=\frac{S_{z}}{2}\cosh\theta \nonumber\\
	K_{0}^{2}&-|K_{+}|^{2}=\frac{S_{z}^{2}}{4},
\end{split}
\end{equation}
so that $\theta$ and $\psi$ parametrize the hyperboloid. The logic behind this choice becomes apparent when we compute the canonical form
\begin{multline}
	\frac{i}{2} \sum_{m}A^*_{m}dA_{m}-A_{m} dA^{*}_{m} = N\, d\chi + S_{z}\,d\phi\\
	+\frac{S_{z}}{2}\cosh\theta\, d\psi,
\end{multline}
where $N$ is the total number of particles $N=\sum_{m}|A_{m}|^{2}$. In this way we can compute the three actions corresponding to the three circuits: $\chi:0\to 2\pi$, $\phi:0\to 2\pi$, and the trajectories on the reduced phase space discussed in Section~\ref{sub:qualitative_features_of_dynamics}. Using Eq.~\eqref{SpinorMonod_actions} we find that the corresponding actions are
\begin{equation}
	\label{SpinorMonod_spinoract}
	\begin{split}
		I_{1}&=N\\
		I_{2}&=S_{z}\\
		I_{3}&=\frac{1}{2\pi}\frac{S_{z}}{2} \oint \cosh\theta\, d\psi
	\end{split}
\end{equation}
The last formula has a geometrical interpretation in terms of the area of the hyperboloid enclosed by the trajectory. This is not the area induced by the usual Euclidean metric, but rather by the `Minkowski' metric
\begin{equation}
	\label{SpinorMonod_metric}
	dK_{z}^{2}-dK_{x}^{2}-dK_{y}^{2}=\left(\frac{S_{z}}{2} \right)^{2}\left(d\theta^{2}-\sinh^{2}\theta\,d\psi^{2}\right),
\end{equation}
(c.f. the quadratic term in Eq.~\eqref{SpinorCondensateMonodromy_HypSpinHam}) with the corresponding area element
\begin{equation}
	\label{SpinorMonod_area}
	dA = \left(\frac{S_{z}}{2} \right)^{2}\sinh\theta\, d\theta\,d\psi
\end{equation}
Thus we have
\begin{equation}
	\label{SpinorMonod_I3area}
	I_{3}=\frac{A}{\pi S_{z}}
\end{equation}
This interpretation will prove useful in the next section when we investigate the behavior of this action.


\subsection{Properties of the action}
\label{sub:action}

\begin{figure}
	\centering
		\includegraphics[width=\columnwidth]{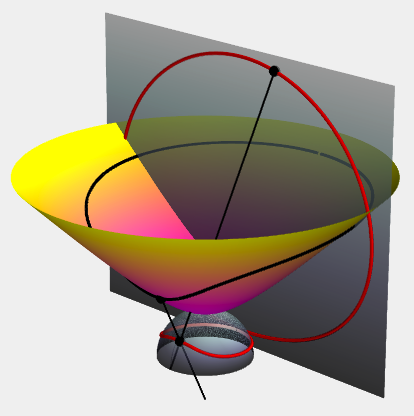}
	\caption{Trajectory on the reduced phase space in the half-plane model. The trajectory is first projected to the hemisphere of radius $s_{z}/2$ from the point $(0,0,-s_{z}/2)$, and from there projected to the plane $k_{x}=-s_{z}/2$ from the point $(s_{z}/2,0,0)$}
	\label{fig:half_plane}
\end{figure}

Let us now investigate the properties of the action $I_{3}$. Note that we are going to continue assuming $S_{z}>0$ to avoid a rash of modulus signs. Due to the symmetry of the Hamiltonian function $h$ in Eq.~\eqref{h_func}, it is convenient to work not with the coordinates $\theta$ and $\psi$ introduced in the preceding section, but instead with the half-plane model of hyperbolic geometry \cite{Cannon:1997}, illustrated in Fig.~\ref{fig:half_plane}. 
The variables on the half plane are
\begin{align}
	\label{half_plane}
\begin{split}
	y = \frac{2k_{y}}{k_{0}-k_{x}}\\
	z = \frac{s_{z}}{k_{0}-k_{z}}.	
\end{split}
\end{align}
In terms of these variables $k_{y}=\frac{s_{z}}{2}\frac{y}{z}$ while
\begin{align}
	\label{kvar}
	\begin{split}
		k_{0}&=\frac{s_{z}}{4}\left[\frac{z}{2}\left(1+\frac{y^{2}}{z^{2}} \right) +\frac{2}{z}\right]\\
		k_{x}&=\frac{s_{z}}{4}\left[\frac{z}{2}\left(1+\frac{y^{2}}{z^{2}} \right)-\frac{2}{z} \right].		
	\end{split}	
\end{align}
The equation relating $y$ and $z$ given $h$ and $s_{z}$ is
%
%
%
\[\left(s_{z}-\tilde q z /2\right)y^{2}=\frac{\tilde q}{2}z^{3}-z^{2}(2h/s_{z})+2z\left(2+\tilde q\right)-4s_{z}\equiv P_{3}(z) \]
so that
\begin{equation}
	\label{ysol}
	y=\pm\sqrt{\frac{P_{3}(z)}{(s_{z}-\tilde q z/2)}}
\end{equation}
%
%
The area element is
\begin{equation}
	\label{SpinorMonod_HalfPlaneAres}
	dA = \left(\frac{S_{z}}{2}\right)^{2}\frac{dy dz}{z^{2}}
\end{equation}
giving the action
\begin{equation}
	\label{newaction}
	I_{3}=\frac{S_{z}}{2\pi}\int_{z_{<}}^{z_{>}} \sqrt{\frac{P_{3}(z)}{(s_{z}-\tilde q z/2)}}\frac{dz}{z^{2}}.
\end{equation}
The endpoints of the integral are two of the positive roots of the cubic $P_{3}(z)$. We discuss the $\tilde q>0$ and $\tilde q<0$ cases separately. 

\subsubsection{$\tilde q<0$} 
\label{ssub:qlessaction}

\begin{figure}
	\centering
		\includegraphics[width=\columnwidth]{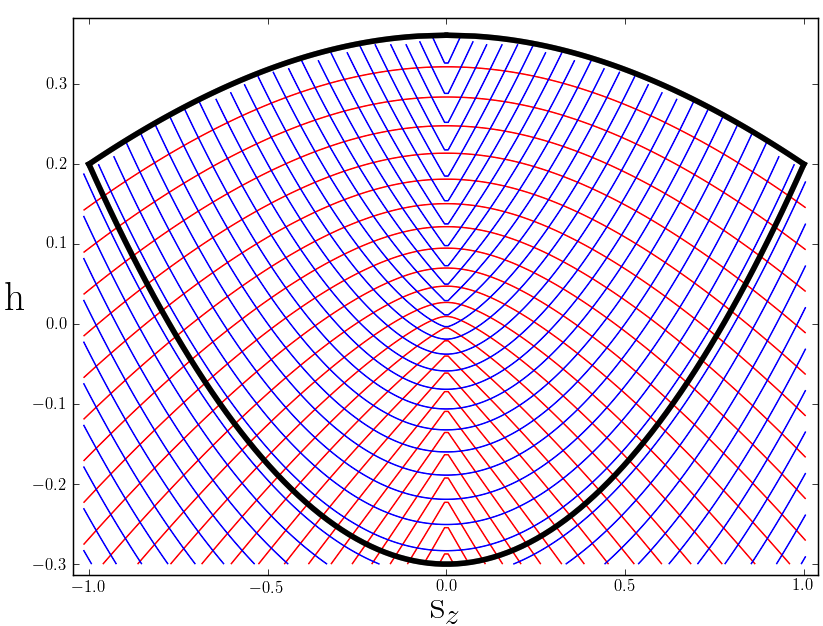}
	\caption{Level sets of the actions $I_{3}/N$ (red) and $I_{3}'=I_{3}/N+|s_{z}|/2$ (blue) for $\tilde q=-0.3$. The first is cusped below $h=0$, the second above.} 
	\label{fig:ActionContours}
\end{figure}

\begin{figure}
\centering
\def\svgwidth{1.5\columnwidth}
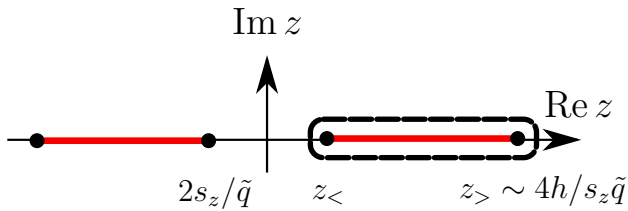
\caption{Branch cuts (red segments) of the integrand and integration contour (dashed) in Eq.~\eqref{newaction} for $\tilde q<0$, $h<0$, and $s_{z}\to 0$. $z_{<}$ and the unmarked root are both $O(s_{z})$.}
\label{fig:contour}
\end{figure}

For $\tilde q<0$ the level sets of the action are shown in Fig.~\ref{fig:ActionContours}. For $h<0$, we see that $I_{3}$ is not smooth along $s_{z}=0$. To understand the origin of this behavior, let us consider the roots of $P_{3}(z)$ as $s_{z}\to 0$. Two roots are positive, and one negative (as should be clear from Fig.~\ref{fig:qlesser}). Further, two of the roots are $O(s_{z})$, being the roots of the quadratic
\[
	2h z^{2}-2s_{z}(2+\tilde q)z+4s_{z}^{2},
\]
while the remaining root is $4h/s_{z}\tilde q$, and diverges to $\pm \infty$ depending on the sign of $h$. For $h>0$ the two positive roots are those that are vanishing. Since these are the limits of the integral, we can see immediately that $I_{3}$ is $O(s_{z}^{0})$. Further, the first correction is at order $s_{z}^{2}$, so $I_{3}$ is smooth about $s_{z}=0$. For $h<0$, one of the positive roots is diverging, corresponding to a trajectory that encircles the tip of the cone. It is this divergence of the upper limit of the integral in Eq.~\eqref{newaction} that is responsible for the cusp. To see this, note that the integrand has four square root singularities to be joined up by two branch cuts. 
For $\tilde q<0$, there is one branch cut on either half of the real axis. The integral for the action is half the integral circulating the branch cut on the positive side (see Fig.~\ref{fig:contour}). We can deform this contour so that it circulates the branch cut on the \emph{negative} side and the pole at $z=0$ -- both giving a contribution of $O(s_{z}^{0})$ -- and the pole at infinity. The contribution of this latter pole gives the cusp
\begin{equation}
	\label{I3lim}
	I_{3}(H,S_{z}) \sim I_{3}(H,0)-\frac{|S_{z}|}{2}\Theta(-H)+O(S_{z}^{2}),
\end{equation}
where we now restore the modulus sign. Since $\partial I_{3}/\partial H<0$ (see Fig.~\ref{fig:qlesser}), this is consistent with the contours in Fig.~\ref{fig:ActionContours}. The implications of the cusp will be discussed in Section~\ref{sub:rotation_number_and_monodromy}. Note that if we define a new action $I_{3}'\equiv I_{3}+|S_{z}|/2$, this has a cusp for $h>0$ instead.


\subsubsection{$\tilde q>0$}
\label{ssub:qgreateraction}

\begin{figure}
	\centering
		\includegraphics[width=\columnwidth]{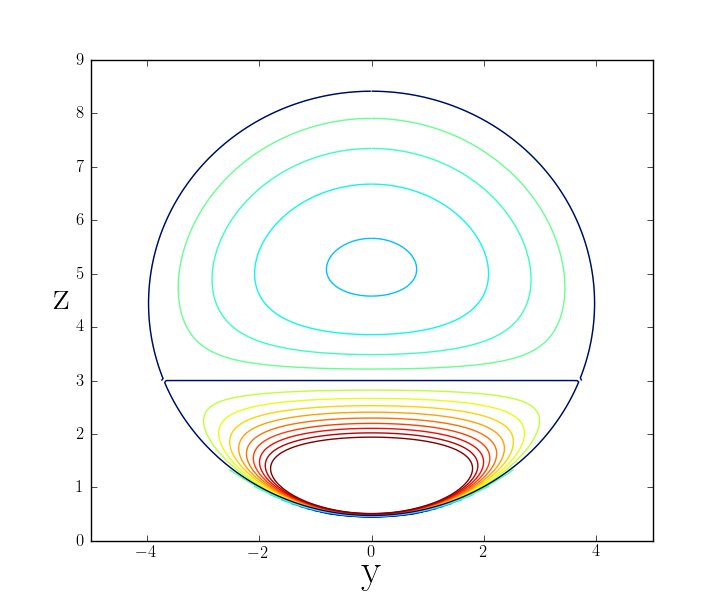}
	\caption{Level sets of $h$ on the half-plane for $s_{z}=0.45$, $\tilde q=0.3$. The dark line is the separatrix corresponding to $h=\frac{1}{2} s_z^{2}+\tilde q$.}
	\label{fig:LobachevskyCircle}
\end{figure}

For $\tilde q >0$ we look for interesting behavior associated with the separatrix at $h=\frac{1}{2}s_{z}^{2}+\tilde q$. In the half plane model the phase space corresponding to the truncated hyperboloid ($k_{0}<1/2$) is a disc bounded by $y^{2}+(z-2/s_{z})^{2}=4(s_{z}^{-2}-1)$ (circles project to circles), and the separatrix consists of this circle plus the chord $z=2s_{z}/\tilde q$ (see Fig.~\ref{fig:LobachevskyCircle}), corresponding to the square root divergence in the action integrand Eq.~\eqref{newaction}.

The roots of $P_{3}(z)$ are all positive (in Fig.~\ref{fig:qgreater} the `third' solution lies on the other branch of the hyperbolas of constant energy, outside of the physical phase space $k_{0}<1/2$). The endpoints $z_{<,>}$ are the greatest two roots at energies below the separatrix and the smallest two above it. These considerations show that the action is discontinuous at the separatrix. Repeating the analysis of the $\tilde q<0$ case shows that below the separatrix the action again develops a cusp
\begin{equation}
	\label{I3lim2}
	I_{3}(H,S_{z}) \sim I_{3}(H,0)-\frac{|S_{z}|}{2}\Theta\left(\frac{1}{2} s_z^{2}+\tilde q-h\right)+O(S_{z}^{2}).
\end{equation}

\subsection{Rotation angle and monodromy}
\label{sub:rotation_number_and_monodromy}

\begin{figure}
\centering
\def\svgwidth{1.0\columnwidth} 
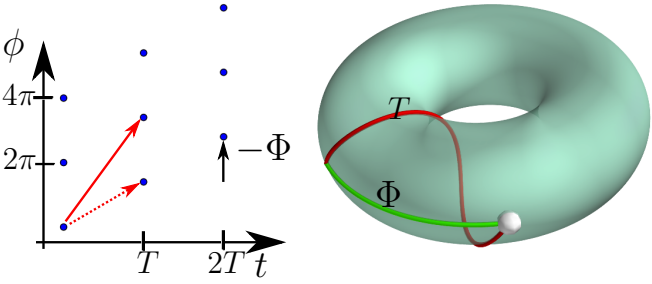
\caption{(Left) As we circle the origin in $S_{z}$, $H$ space for $\tilde q<0$ the period lattice is deformed continuously, returning to its original form, but after shifting the lattice vector corresponding to $I_{3}$ by $2\pi$ in the $\phi$ direction. (Right) Schematic illustration of the rotation angle. While executing a single period $T$ of motion on the reduced phase space the system rotates by an angle $\Phi$}
\label{fig:PeriodLattice}
\end{figure}

More significant than the action is its derivatives, which give us the period lattice vectors from Eq.~\eqref{SpinorMonod_PeriodLattice}. Specializing to the actions of Eq.~\eqref{SpinorMonod_spinoract} we find
\begin{equation}
	\label{SpinorMonod_PeriodLatticeSpecial}
	(\mathbf{e}_{i})_{j}=2\pi\begin{pmatrix}
		1 & 0 & 0 \\
		0 & 1 & 0 \\
		\frac{\partial I_{3}}{\partial N} & \frac{\partial I_{3}}{\partial S_{z}} & \frac{\partial I_{3}}{\partial H} 
	\end{pmatrix}
\end{equation}
The vector $\mathbf{e}_{3}=2\pi\begin{pmatrix}
	\frac{\partial I_{3}}{\partial N} & \frac{\partial I_{3}}{\partial S_{z}} & \frac{\partial I_{3}}{\partial H} 
\end{pmatrix}$ tells us how to execute a closed orbit around the third circle of the three-torus: we evolve for a time $2\pi \frac{\partial I_{3}}{\partial H}$ (this is then the period of the motion on the reduced phase space), change the overall phase of the spinor by $2\pi \frac{\partial I_{3}}{\partial N}$, and rotate about the $z$-axis by $2\pi \frac{\partial I_{3}}{\partial S_{z}}$. The \emph{rotation angle} 
\begin{equation}
	\label{SpinorMonod_RotationDef}
	\Phi(S_{z},H) \equiv -2\pi \frac{\partial I_{3}}{\partial S_{z}}
\end{equation}
is therefore the rotation about the $z$-axis associated with one period of the reduced motion (see Fig.~\ref{fig:PeriodLattice}). Comparing with Eq.~\eqref{I3lim} we arrive at the surprising conclusion that for $\tilde q<0$, $\Phi(S_{z},H)$ is not a single-valued function, but rather increases by $2\pi$ upon encircling the origin $H=S_{z}=0$. By contrast the period 
\begin{equation}
	\label{SpinorMonod_Period}
	T \equiv 2\pi\frac{\partial I_{3}}{\partial H}
\end{equation}
is single-valued (though logarithmically diverging as we pass through the origin). $T$ may be expressed as an elliptic integral \cite{Zhang:2005}. Note that for $\tilde q>0$ the separatrix divides the phase space into two disjoint regions (see Fig.~\ref{fig:LobachevskyCircle}). In each of these regions action-angle coordinates can be introduced without difficulty.

The non-trivial mapping of the period lattice into itself upon encircling the origin in $(S_{z},H)$ space is the characteristic signature of \emph{monodromy} (see Fig.~\ref{fig:PeriodLattice}), and by Eq.~\eqref{SpinorMonod_angles} corresponds to angle variables that are not single-valued. The mapping of the period lattice vectors is written as
\begin{equation}
	\label{SpinorMonod_PeriodMap}
	\mathbf{e}_{i}\to \mathbf{e}_{i}'=M_{ij}\mathbf{e}_{j},
\end{equation}
where $M$ is a integer-valued matrix of unit determinant (an element of the group $SL(3,\mathbb{Z})$) called the monodromy matrix. In our case
\begin{equation}
	\label{SpinorMonod_MonodMat}
	M = \begin{pmatrix}
		1 & 0 & 0 \\
		0 & 1 & 0 \\
		0 & 1 & 1
	\end{pmatrix}
\end{equation}
Note that we can focus on the $2-3$ subspace of the period lattice. Nothing interesting happens in the direction corresponding to $I_{1}=N$, reflecting the fact that $N$ was scaled out of the problem.

What is special about the point $S_{z}=H=0$ for $\tilde q<0$? Recall from the discussion of Section \ref{sub:qualitative_features_of_dynamics} that for $S_{z}=0$ the reduced phase space is a cone and for $H<0$ the trajectory encircles the tip, while for $H>0$ it does not. From Eq.~\eqref{SpinorMonod_spinorparam} the tip of the cone corresponds to the state $(a_{1},a_{0},a_{-1})=(0,e^{-i\chi},0)$, which is invariant under rotations about the $z$-axis. Thus the torus is \emph{pinched} at this point: the circle corresponding to rotations through $\phi$ has contracted to nothing. Without such a singularity there would be no distinction between paths that circuit the origin and those that do not, and hence no possibility of non-trivial monodromy. Further, the structure of the singularity -- known as a \emph{focus-focus singularity} in the mathematical literature -- actually fixes the monodromy without the need for explicit calculation of the actions \cite{zou1992,zung1997,cushman2001}. To see this, let us consider the quadratic Hamiltonian in the vicinity of the tip of the cone. After fixing $a_{0}=1$, Eq.~\eqref{SpinorCondensateMonodromy_SimpleHam} gives
\begin{equation}
	\label{SpinorMonod_Quad}
	h_{\text{quad}}=|a^{\vphantom{*}}_{1}+a_{-1}^{*}|^{2}+\tilde q\left(|a_{1}|^{2}+|a_{-1}|^{2}\right).
\end{equation}
In the range $-2<\tilde q<0$ this corresponds to an `inverted' complex oscillator, as may be seen by defining
\begin{equation}
	\label{SpinorMonod_CompOsc}
\begin{split}
	z \equiv \frac{1}{\sqrt{2}}\left(\frac{2-|\tilde q|}{|\tilde q|}\right)^{1/4}\left(a_{1}+a_{-1}^{*}\right)\\
	\varpi \equiv -\frac{i}{\sqrt{2}}\left(\frac{|\tilde q|}{2-|\tilde q|}\right)^{1/4}\left(a_{1}-a_{-1}^{*}\right).
\end{split}
\end{equation}
with $\left\{z,\varpi^{*}\right\}_{\text{PB}}=1$. In terms of these variables 
\begin{equation}
	\label{SpinorMonod_QuadComp}
	h_{\text{quad}}=\Omega\left(|z|^{2}-|\varpi|^{2}\right)
\end{equation}
where $\Omega = \sqrt{|\tilde q|(2-|\tilde q|)}$. The unstable and stable modes are then
\begin{equation}
	\label{SpinorMonod_StableUnstable}
	a_{\pm}\equiv \frac{1}{\sqrt{2}}\left(z\mp \varpi\right)
\end{equation}
satisfying $\left\{a_{+},a_{-}^{*}\right\}_{\text{PB}}=1$. In terms of these modes
\begin{align}
	\label{SpinorMonod_hszmodes}
\begin{split}
	h &= \Omega\left(a_{+}^{*}a_{-}+a_{-}^{*}a_{+}\right)\\
	s_{z} &= a_{-}^{*}a_{+}-a_{+}^{*}a_{+}.
\end{split}
\end{align}
The linearized equations of motion are $\dot a_{\pm}=\pm\Omega a_{\pm}$, showing that as $a_{+}$ grows exponentially, $a_{-}$ decays so as to conserve $a_{+}^{*}a_{-}$. We now recapitulate an argument from Ref.~\cite{babelon:2009} that shows how these simple considerations fix the monodromy.

\begin{figure}
\centering
\def\svgwidth{1\columnwidth}
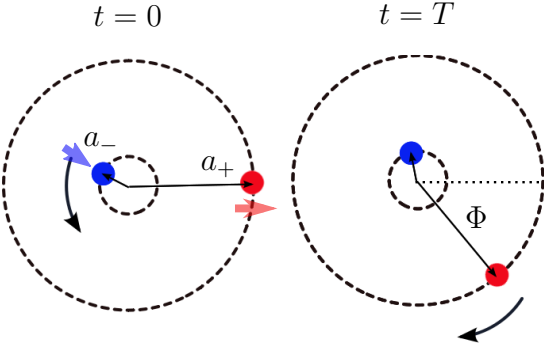
\caption{After one period of motion on the reduced phase space the system returns close to the pinch, but rotates by the rotation angle $\Phi$. The angle between $a_{+}$ and $a_{-}$ remains constant. We can circle the origin $h=s_{z}=0$ by winding $a_{-}(0)$ as shown, keeping $a_{+}(0)$ fixed. By the argument in the text, $a_{+}(T)$ must wind in the opposite direction, showing that the rotation angle changes by $2\pi$.}
\label{fig:StableUnstable}
\end{figure}

From Eq.~\eqref{SpinorMonod_hszmodes}, varying the overall phase of $a_{+}^{*}a_{-}$ amounts to circling the origin $h=s_{z}=0$. When either $a_{+}$ or $a_{-}$ vanishes (and the other is small), we are on the pinched torus, the two components corresponding to the stable and unstable branches respectively. These two branches are of course connected away from the linear regime. If we start from $|a_{-}(0)|\ll |a_{+}(0)|=\eta$ at time $t=0$ for some $\eta\ll 1$ the system is close to the unstable branch of the pinched torus and will evolve in finite time $t'<T$ to be close to the stable branch i.e. $|a_{+}(t')|\ll |a_{-}(t')|=\eta$. The key observation is that the limit $a_{-}(0)\to 0$ is well behaved: we approach the pinched torus but the evolution $0\to t'$ excludes the `pinch'. Thus the phase of $a_{-}(t')$ will not wind with the phase of $a_{-}(0)$. However, since $a_{+}^{*}a_{-}$ is conserved, $a_{+}(t')$ must wind \emph{oppositely} to the phase of $a_{-}(0)$. After evolving for an additional time $\Omega^{-1}\ln|\frac{a_{+}(0)}{a_{-}(0)}|$ to give a total time of one period $T$ of the reduced motion, $|a_{+}|$ grows back to its initial value $\eta$, and $|a_{-}|$ decays to its initial value, as the system passes close to the pinch. The phases of $a_{\pm}$ will have changed, however, and this change is just the rotation angle (as should be clear from the definitions in  Eq.~\eqref{SpinorMonod_CompOsc}). The winding of $a_{+}(T)$ in the opposite sense to the winding of $a_{-}(0)$ corresponds to a $2\pi$ change of the rotation angle, see Fig.~\ref{fig:StableUnstable}.

\begin{figure*}
	\centering		\includegraphics[width=\columnwidth]{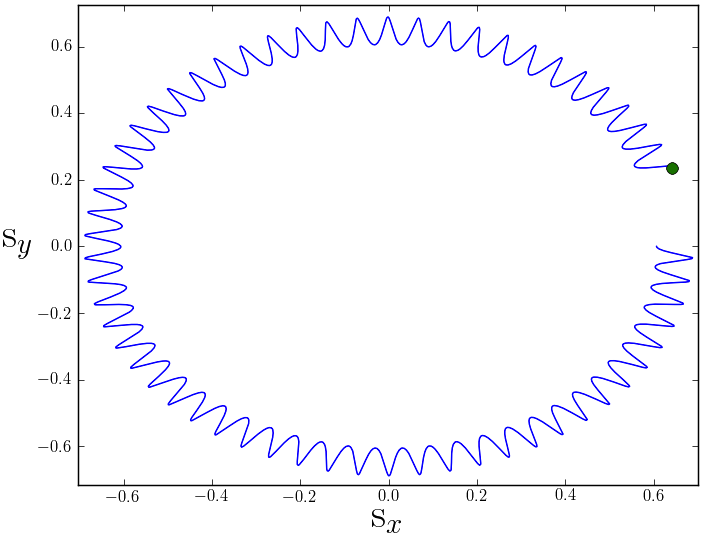}\includegraphics[width=\columnwidth]{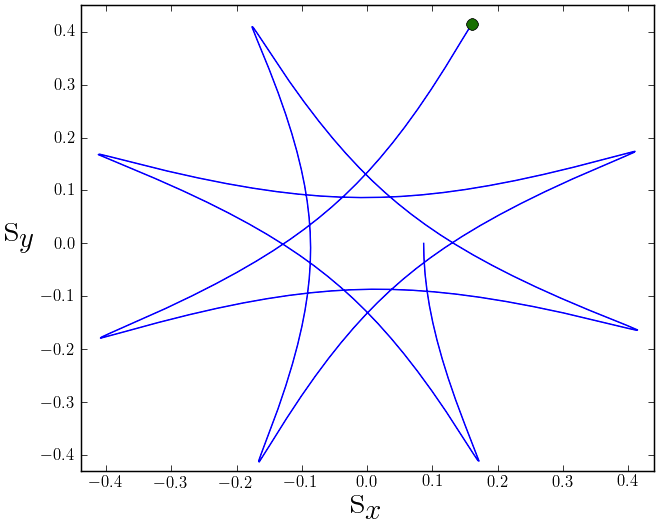}
	\caption{Evolution of transverse magnetization for $\tilde q=-0.3$. (Left) Initial conditions $a_{1}=0.4$, $a_{-1}=0.3$ (Right) Initial conditions $a_{1}=0.4$, $a_{-1}=-0.3$, rotation angle approaching $+\pi$}
	\label{fig:RotationLarge}
\end{figure*}

We close this section by giving a more explicit illustration of the rotation angle. If we fix the $a_{0}$ component of the spinor to be real, so that $a_{0}=\sqrt{1-|a_{1}|^{2}-|a_{-1}|^{2}}$, the Gross--Pitaevskii equations become
\begin{widetext}
	\begin{equation}
		\label{SpinorCondensateMonodromy_diffmotion}
		i\dot a_{\pm 1}=\left[\pm s_{z}a_{\pm 1}+(a^{*}_{\mp 1}+a_{\pm 1})(1-|a_{1}|^{2}-|a_{-1}|^{2})-(a^{*}_{1}a^{*}_{-1}+a_{1}a_{-1}+|a_{1}|^{2}+|a_{-1}|^{2})a_{\pm 1}\right]+\tilde q a_{\pm 1}.
	\end{equation}	
\end{widetext}
(we measure time in units of $e_{2}^{-1}$). The transverse magnetization $s_{+}=s_{x}+is_{y} =\sqrt{2}(a^{*}_{1}+a_{-1})\sqrt{1-|a_{1}|^{2}-|a_{-1}|^{2}}$. The evolution of $s_{+}$ for small and large (approaching $\pi$) rotation angles is shown in Fig.~\ref{fig:RotationLarge}.

\subsection{Semiclassical quantization} 
\label{sub:semiclassical_quantization}

\begin{figure}
	\centering
		\includegraphics[width=\columnwidth]{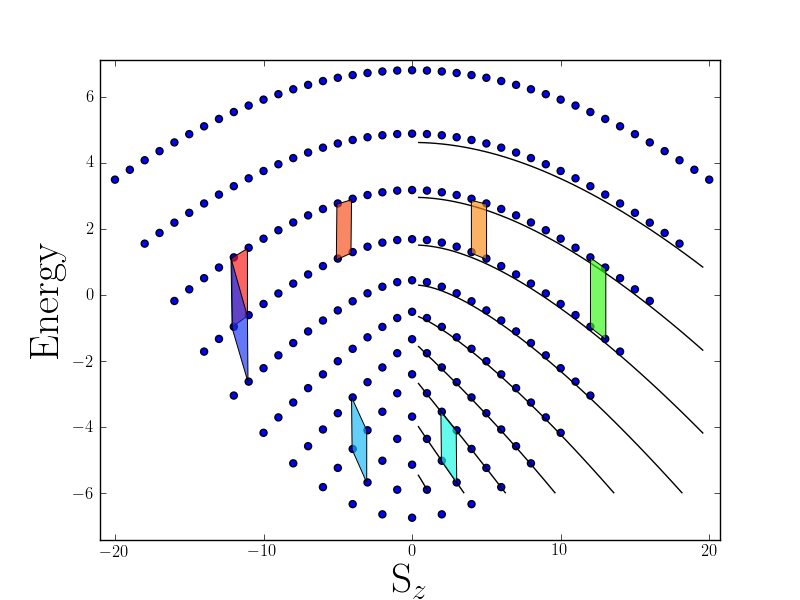}
	\caption{Spectrum of Eq.~\eqref{SpinorCondensateMonodromy_SimpleHam} with $N=20$ particles and $\tilde q=-0.3$. The dark lines are the integral contours of the action $I_{3}$. Transporting an elementary cell of the lattice around the origin leads to non-trivial monodromy.}
	\label{fig:action_contours}
\end{figure}

With this extensive groundwork laid we can finally discuss the quantization of the problem. The semiclassical prescription of Einstein, Brillouin and Keller (EBK) is to quantize the actions according to 
\begin{equation} 
	\label{SpinorMonod_EBK}
	I_{i}=\left(n_{i}+\mu_{i}\right)\hbar,\qquad n_{i}\in\mathbb{Z}
\end{equation}
where the $\mu_{i}$ are known as \emph{Maslov indices}. In the present case there are no non-trivial Maslov indices, and we have the quantization rule (recalling that we set $\hbar=1$ in writing down the actions in Section~\ref{sub:general_consequences_of_integrability})
\begin{equation}
	\label{SpinorMonod_QuantRules}
	I_{3}=\frac{1}{2\pi}\frac{S_{z}}{2} \oint \cosh\theta\, d\psi \in \mathbb{Z}_{+}
\end{equation}
(the other two rules simply quantize $N$ and $S_{z}$ in the familiar way). The integral is related to the hyperbolic area enclosed as explained earlier. The integer contours of the action are shown in Fig.~\ref{fig:action_contours}, compared with the result of numerically diagonalizing the Hamiltonian Eq.~\eqref{SpinorCondensateMonodromy_SimpleHam}. A more accurate semiclassical quantization could be achieved using the results of Ref.~\cite{kurchan:1989}, but we do not require it here.

Above the origin the eye naturally picks out negatively curved rows in the spectrum, and these correspond to the contours of the action $I_{3}$. Beneath the origin these rows are not so evident, reflecting the fact that the $I_{3}$ is not smooth here. Instead the eye picks out positively curved rows corresponding to the action $I_{3}'=I_{3}+|S_{z}|/2$.

An elegant description of the relation between the quantum spectrum in the semiclassical limit and the monodromy of the classical system has been given by Zhilinskii \cite{zhilinskii:2006}. The EBK quantization rules tells us that locally the eigenstates form a lattice in the space of conserved quantities $F_{j}$ $j=1,\ldots N$ with lattice vectors given by changing each of the $I_{i}$ by $\hbar$
\begin{equation}
	\label{SpinorMonod_QuantumLatticeVectors}
	\hbar \left(\frac{\partial F_{j}}{\partial I_{i}}\right)_{I_{k}\neq I_{i}\text{ fixed}}=\hbar \left(\bm{\epsilon_{i}}\right)_{j}
\end{equation}
(c.f. Eq.~\eqref{SpinorMonod_AngVelMatrix}). Thus the lattice of quantum states is just the reciprocal lattice of the period lattice. Accompanying the mapping of the period lattice vectors on circling the origin in Eq.~\eqref{SpinorMonod_PeriodMap} is the corresponding map on the `quantum' lattice
\begin{equation}
	\label{SpinorMonod_QuantumMap}
	\bm{\epsilon_{i}}\to \bm{\epsilon}_{i}'=\left(M^{-1}\right)_{ji}\bm{\epsilon}_{j},
\end{equation}
with
\begin{equation}
	\label{SpinorMonod_QuantumMonodMat}
	\left(M^{-1}\right)^T = \begin{pmatrix}
		1 & 0 & 0 \\
		0 & 1 & -1 \\
		0 & 0 & 1
	\end{pmatrix}
\end{equation}
The resulting `defect' is illustrated in Fig.~\ref{fig:action_contours}.

These semiclassical considerations will be put on firmer footing in the next section, where we discuss the exact solution of the quantum problem.



\section{Solution of the quantum Hamiltonian} 
\label{sec:solution_of_the_quantum_hamiltonian}

\subsection{The Bethe ansatz equations}
\label{sub:the_bethe_ansatz_equations}

In Ref.~\cite{Bogoliubov:2006}, Bogoliubov used the algebraic Bethe ansatz (ABA) to solve the Hamiltonian Eq.~\eqref{SpinorCondensateMonodromy_SimpleHam}, based on the mapping to hyperbolic spins \footnote{The Hamiltonian in Eq.~\eqref{SpinorCondensateMonodromy_HypSpinHam} has the form of a Richardson--Gaudin model, originally studied for $SU(2)$ spins \cite{dukelsky2004}.  The hyperbolic case has also been studied repeatedly in a variety of contexts, see e.g. Refs.~\cite{dukelsky2001,balantekin2004,ovchinnikov2004}}. An eigenstate of the Hamiltonian is written as
\begin{equation}
	\label{SpinorMonod_BAcreate}
	\ket{\Psi}=\prod_{j}^{N_{R}}\left(\frac{B^{+}}{\lambda_{j}}+\frac{K^{+}}{\lambda_{j}+N\tilde q/2} \right)\ket{0,\nu_{K}}_{K}\otimes\ket{0,\nu_{B}}_{B}
\end{equation}
where the $\lambda_{j}$ $j=1,\ldots N_{R}$ satisfy the equations
\begin{equation}
	\label{SpinorCondensateMonodromy_BAEqs}
	1- \frac{\nu_{B}}{\lambda_{j}}-\frac{\nu_{K}}{\lambda_{j}+N\tilde q/2}=\sum_{l\neq j}^{N_{R}} \frac{1}{\lambda_{j}-\lambda_{l}}.
\end{equation}
Recall that $\nu_{K}=\frac{1}{2}\left(|S_{z}|+1\right)$ and $\nu_{B}=\frac{1}{4},\frac{3}{4}$ according to the parity of $N-S_{z}$ (see Eq.~\eqref{SpinorMonod_parity}). Since each factor in Eq.~\eqref{SpinorMonod_BAcreate} creates a pair of particles the total number $N$ is related to the number of roots $N_{R}$ by
\begin{equation}
	\label{SpinorMonod_NNroots}
	N = 2N_{R}+|S_{z}|+2\nu_{B}-\frac{1}{2}
\end{equation}
The energy of the state Eq.~\eqref{SpinorMonod_BAcreate} is
\begin{equation}
	\label{SpinorMonod_BAenergy}
	E\left(\left\{\lambda_{j}\right\}\right)=\frac{e_{2}}{2N}N(N-1)+q|S_{z}|-\frac{4e_{2}}{N}\sum_{j=1}^{N_{R}}\lambda_{j}
\end{equation}
(The more complicated form in Ref.~\cite{Bogoliubov:2006} is expressed in terms of $N_{R}$)

In the next section we will give a derivation of these results that does not rely on the full machinery of the ABA and is suited to semiclassical approximations. For the remainder of this section we discuss the solutions of Eq.~\eqref{SpinorCondensateMonodromy_BAEqs} in qualitative terms.

The simplest way to understand the character of the solutions is to interpret Eq.~\eqref{SpinorCondensateMonodromy_BAEqs} as the extremal condition of the `potential'
\begin{equation}
	\label{SpinorCondensateMonodromy_FictPot}
	\sum_{i<j}\ln|\lambda_{i}-\lambda_{j}|+\sum_{i}\left[\nu_{B}\ln|\lambda_{i}|+\nu_{K}\ln|\lambda_{i}+N\tilde q/2|-\lambda_{i}\right].
\end{equation}
This corresponds to a set of $N_{R}$ unit positive charges located at positions $\left\{\lambda_{i}\right\}$ interacting among themselves a 2D Coulomb potential, with a pair of positive charges fixed at $0$ and $-N\tilde q/2$ with strengths $\nu_{B}$ and $\nu_{K}$ respectively. Additionally there is a constant electric field pushing the charges in a negative direction. Note that the positions of the charges $\lambda_{i}$ \emph{could} be complex numbers, as in the related systems of equations considered in Ref.~\onlinecite{Shastry:2001}, for instance. Since all charges in the present case are of the same sign, however, it is not hard to see that equilibrium configurations can only occur for all $\lambda_{i}$ real.

Let's first consider the special case $\tilde q=0$, which corresponds to zero magnetic field. In this case the spectrum is simple, consisting of $SU(2)$ multiplets. This can be readily understood in terms of the above equations, where we now have a single fixed charge of strength $\nu_{B}+\nu_{K}$ at the origin. Let's start from a solution with given $S_{z}$. We can generate another with $S_{z}-2$, by adding one root (this keeps the number of particles fixed, see Eq.~\eqref{SpinorMonod_NNroots}) at the origin. The charge of the fixed charge decreases by one unit, but this is compensated by the new root, so the same configuration of the other roots is still a solution. Because the new root is at the origin, the total energy is unchanged (see Eq.~\eqref{SpinorMonod_BAenergy}). Proceeding in this way, and using both values of $\nu_{B}$, we can generate the whole $SU(2)$ multiplet. 

It is straightforward to check that the correct energies are reproduced in this case. Multiplying the Bethe ansatz equations Eq.~\eqref{SpinorCondensateMonodromy_BAEqs} by $\lambda_{j}$ and summing over $j$ gives (assuming that there are no charges at the origin)
\begin{equation}
	\label{SpinorMonod_SU2case}
	\sum_{j=1}^{N_{R}}\lambda_{j}=\frac{1}{2}N(N+2\nu_{B}+|S_{z}|)
\end{equation}
giving the energy
\begin{equation}
	\label{SpinorMonod_SU2energy}
	E=\frac{e_{2}}{2N}\begin{cases}
		S_{z}(S_{z}+1)-2N, &\text{ if }N-S_{z}\text{ even} \\
		(S_{z}+1)(S_{z}+2)-2N, &\text{ if }N-S_{z}\text{ odd} \\
	\end{cases}
\end{equation}
Which are the eigenvalues of $\frac{e_{2}}{2N}: \hat{\mathbf{S}}\cdot\hat{\mathbf{S}}:=\frac{e_{2}}{2N}\left( \hat{\mathbf{S}}\cdot\hat{\mathbf{S}}-2N\right)$ for total spin $S=S_{z}$ and $S=S_{z}+1$ respectively. Note that Bose statistics limits $S$ to even values for $N$ even and odd values for $N$ odd.

At finite $\tilde q$ the two fixed charges separate. When adding a root, reducing $S_{z}$ by 2, one can choose to place it either between the fixed charges or in the region $\lambda >\max(-N\tilde q/2,0)$ (some rearrangement of the other charges occurs). These two moves may be used to build up the spectrum and correspond to increments in two different choices for the action in semiclassical quantization (see Fig.~\ref{fig:action_contours}). Moving between consecutive blue contours while staying on the same red contour corresponds to adding roots between the charges, and these go over to the $SU(2)$ multiplets as $q\to 0$. Conversely moving between consecutive red contours while staying on the same blue contour corresponds to adding roots in the region $\lambda >\max(-N\tilde q/2,0)$. In Section~\ref{sub:wkb_analysis_of_the_schr"odginer_equation} we will verify that the cusps in these actions are reproduced correctly.

\subsection{Derivation of the Bethe ansatz equations} 
\label{sub:derivation_of_the_bethe_ansatz_equations}

Let us derive Eq.~\eqref{SpinorCondensateMonodromy_BAEqs} without employing the full machinery of the ABA. We begin by finding the discrete Schr\"odinger equation for the operator
\[\mathcal{H}=N\tilde q K^{0}+2B^{0}K^{0}-B^{+}K^{-}-B^{-}K^{+}.\]
Writing an eigenstate as
\begin{equation}
	\label{SpinorCondensateMonodromy_eigenstate}
	\ket{\Psi}=\sum_{n=0}^{N_{R}}c_{n}\left(B^{+}\right)^{N_{R}-n}\left(K^{+}\right)^{n}\ket{\nu_{K},\nu_{B}},	
\end{equation}
we obtain the following equation for the coefficients $c_{n}$
\begin{multline}
	\label{SpinorCondensateMonodromy_discrete}
	\left[N\tilde q(n+\nu_{K})+2(N_{R}-n+\nu_{B})(n+\nu_{K})\right]c_{n}\\
	-(N_{R}-n+1+\nu_{B})(N_{R}-n+\nu_{B})c_{n-1}\\
	-(n+1)(n+2\nu_{K})c_{n+1}=\xi c_{n}
\end{multline}
for a state with eigenvalue $\xi$. Now we going to recast the problem as a differential equation for the polynomial
\begin{subequations}
	\label{SpinorCondensateMonodromy_wavefunction}
	\begin{align}
	\Psi(\lambda)&=\sum_{n=0}^{N_{R}}c_{n}(-1)^{n}\lambda^{N_{R}-n+\nu_{B}}\left(\lambda+N\tilde q/2\right)^{n+\nu_{K}}\label{wave1}\\
	&\propto\lambda^{\nu_{B}}\left(\lambda+N\tilde q/2\right)^{\nu_{K}}\prod_{n}\left(\lambda-\lambda_{n}\right).\label{wave2}
	\end{align}
\end{subequations}
With some lengthy algebra one can show that the discrete Eq.~\eqref{SpinorCondensateMonodromy_discrete} is equivalent to the following differential equation for Eq.~\eqref{wave1}
\begin{widetext}
\begin{equation}
	\label{SpinorCondensateMonodromy_Schrodinger}
	\Psi''-2\Psi'-\left[\frac{\nu_{K}(\nu_{K}-1)}{\lambda^{2}}+\frac{\nu_{B}(\nu_{B}-1)}{\left(\lambda+N\tilde q/2\right)^{2}}-\frac{2(N_{R}+\nu_{K}+\nu_{B})\left(\lambda+N\tilde q/2\right)}{\lambda(\lambda+N\tilde q/2)}\right]\Psi=\frac{\xi\Psi}{\lambda(\lambda+N\tilde q/2)}
\end{equation}
Now using Eq.~\eqref{wave2} we can check that if Eqs.~\eqref{SpinorCondensateMonodromy_BAEqs} are satisfied we have a solution to Eq.~\eqref{SpinorCondensateMonodromy_Schrodinger} with eigenvalue 
\begin{equation}
	\label{SpinorCondensateMonodromy_energyresult}
	\xi = 2\nu_{B}\nu_{K}+N\tilde q\nu_{K}\left(1+\sum_{n}\frac{1}{\lambda_{n}+N\tilde q/2} \right).
\end{equation}
%
The relation between second order linear differential equations and equations of Bethe ansatz type is known as the \emph{Heine--Stieltjes} problem. Finally we can use the Bethe equations again to show
\begin{equation}
	\label{SpinorMonod_BetheEnergy}
	-2N\tilde q\nu_{K}\sum_{n}\frac{1}{\lambda_{n}+N\tilde q/2}=\sum_n \lambda_{n}-\frac{N_{R}(N_{R}-1)}{2}-N_{R}(\nu_{K}+\nu_{B}).	
\end{equation}
After restoring all factors, plus the c-number pieces that we have dropped in passing to hyperbolic spins, we obtain the eigenenergy Eq.~\eqref{SpinorMonod_BAenergy}.

For the purposes of semiclassical analysis, it is convenient to recast Eq.~\eqref{SpinorCondensateMonodromy_Schrodinger} as a conventional Schr\"odinger equation using the transformation $\chi=\Psi e^{-\lambda}$ so that $\chi''-\chi=\left(\Psi''-2\Psi'\right)e^{-\lambda}$, giving
\begin{equation}
	\label{SpinorCondensateMonodromy_New_Sch}
	-\chi''+t(\lambda)\chi=0,
\end{equation}
with
\begin{equation}
	\label{SpinorCondensateMonodromy_tevalue}
\begin{split}
	t(\lambda)&=1+\frac{\nu_{B}(\nu_{B}-1)}{\lambda^{2}}+\frac{\nu_{K}(\nu_{K}-1)}{(\lambda+N\tilde q/2)^{2}}-\left(\frac{1}{\lambda}+\frac{1}{\lambda+N\tilde q/2} \right)\overbrace{\left(N_{R}+\nu_{B}+\nu_{K}\right)}^{\text{eigenvalue of }B^{0}+K^{0}}\\
	&\qquad+\frac{2}{N\tilde q} \left(\frac{1}{\lambda}-\frac{1}{\lambda+N\tilde q/2} \right)\overbrace{\left(\xi-N\tilde q\left( N_{R}+\nu_{B}+\nu_{K}\right)/2\right)}^{\text{eigenvalue of }\mathcal{H}-N\tilde q(B^{0}+K^{0})/2}.\\
	&=\frac{P_{4}(\lambda)}{\lambda^{2}(\lambda+N\tilde q/2)^{2}},
\end{split}
\end{equation}
\end{widetext}
where $P_{4}(\lambda)$ is a fourth order polynomial. Notice that $t(\lambda)$ is just the eigenvalue of the transfer operator introduced in Ref.~\cite{Bogoliubov:2006}. The resulting potential describes motion with combined `centrifugal' and Coulomb potentials, the latter being of variable sign.

The derivation of a Schr\"odinger equation for a related system was given in Ref.~\cite{EnolSkii:1993} using the method of separation of variables, but we shall not elaborate on this connection here.


\subsection{WKB analysis of the Schr\"odinger equation} 
\label{sub:wkb_analysis_of_the_schr"odginer_equation}

An understanding of the spectrum for $N$ large requires a semiclassical analysis of Eq.~\eqref{SpinorCondensateMonodromy_New_Sch} by the WKB method, according to which the actions
\begin{equation}
	\label{SpinorMonod_WKBact}
	I_{a,b}=\frac{1}{2\pi}\oint_{\mathcal{R}_{a,b}} d\lambda\,\sqrt{-t(\lambda)}=n_{a,b}+\frac{1}{2},
\end{equation}
with $n_{a,b}\in \mathbb{Z}_{+}$. When $\nu_{B,K}$ are not large the \emph{Langer modification} $\nu_{B,K}(\nu_{B,K}-1)\to \left(\nu_{B,K}-\frac{1}{2}\right)^{2}$ in Eq.~\eqref{SpinorCondensateMonodromy_tevalue} is required, as the naive WKB wavefunction behaves as 
\begin{equation}
	\label{SpinorMonod_WKBlims}
	\chi(\lambda)\propto \begin{cases}
		\lambda^{\frac{1}{2}+\left[\nu_{B}(\nu_{B}-1)\right]^{1/2}}, &\text{ when }\lambda\to 0\\
		(\lambda+N\tilde q/2)^{\frac{1}{2}+\left[\nu_{K}(\nu_{K}-1)\right]^{1/2}}, &\text{ when }\lambda\to -N\tilde q/2\\		
	\end{cases}
\end{equation}
whereas the correct exponents are $\nu_{B}$ and $\nu_{K}$ respectively (see Eq.~\eqref{SpinorCondensateMonodromy_wavefunction}).

The two actions $I_{a,b}$ correspond to motion between the two endpoints (roots of $P_{4}(\lambda)$) contained in the regions
\begin{equation}
	\label{SpinorMonod_regiona}
	\begin{split}
	\mathcal{R}_{a}&:\min(0,-N\tilde q/2)<\lambda<\max(0,-N\tilde q/2)\\		
	\mathcal{R}_{b}&:\max(0,-N\tilde q/2)<\lambda<\infty.
	\end{split}
\end{equation}
$\frac{1}{\pi}\sqrt{-t(\lambda)}$ is the root density of the Bethe roots in the $N\to\infty$ limit. It is possible to obtain the root density directly from the Bethe ansatz equations without using the Schr\"odinger equation (see Appendix~\ref{sec:bethecont}).

The above quantization conditions amount to fixing integer numbers of roots in the two regions. From Eq.~\eqref{SpinorMonod_NNroots} for the total number of roots, we have $I_{b}=-I_{a}-|S_{z}|/2+\text{const.}$, so that if $I_{a}$ is smooth as $S_{z}\to 0$, then $I_{b}$ is not, and vice versa.  
 
To understand how the action can have a cusp, consider the $\tilde q<0$ case, in which case $-N\tilde q/2$ separates $\mathcal{R}_{a}$ and $\mathcal{R}_{b}$. As $S_{z}\to 0$ the centrifugal potential at $-N\tilde q/2$ is vanishing, leaving the `Coulomb' part. For $\xi>0$ ($\xi<0$) this Coulomb potential is repulsive (attractive) for $\lambda>-N\tilde q/2$ and attractive (repulsive) for $\lambda<-N\tilde q/2$. It is then not hard to show that the action corresponding to the region with the attractive potential has a cusp, as the turning point of the action `falls in' to $\lambda=-N\tilde q/2$ as $S_{z}\to 0$. In the vicinity of this point the integral Eq.~\eqref{SpinorMonod_WKBact} looks like
\begin{multline}
	\label{SpinorCondensateMonodromy_gamma_action}
	\frac{1}{\pi}\int d\lambda\sqrt{\frac{2}{N\tilde q}\frac{\xi}{\lambda+N\tilde q/2}-\frac{S_{z}^{2}}{4(\lambda+N\tilde q/2)^{2}}}\\=\frac{1}{\pi}\int_{S_{z}} du\, \sqrt{1-\frac{S_{z}^{2}}{u^{2}}}
\end{multline}
where $u=\sqrt{\frac{2\xi(\lambda+N\tilde q/2)}{N\tilde q}}$ (note that we are concerned with $S_{z}$ of $O(N)$). This is appropriate to the side on which the turning point is falling in, because then the contributions from the other parts of the potential can be ignored. To find the contribution to the integral consider the derivative
\[
		\frac{d}{dS_{z}}\int_{S_{z}} du\, \sqrt{1-\frac{S_{z}^{2}}{u^{2}}}=-S_{z}\int_{S_{z}} \frac{du}{u} \frac{1}{\sqrt{u^{2}-S_{z}^{2}}}=-\frac{\pi}{2}
\]
Thus the corresponding action behaves as $-|S_{z}|/2$ as $S_{z}\to 0$: $I_{a}$ has a cusp for $E>0$ and $I_{b}$ is cusped for $E<0$. The discussion at the end of Section~\ref{sub:the_bethe_ansatz_equations} can then be sharpened with the identification (up to constants) $I_{3}=I_{b}$, $I_{a}=-I_{3}'=-I_{3}-|S_{z}|/2$. In this way we recover the monodromy discussed in Section~\ref{sub:rotation_number_and_monodromy}.



\section{Conclusion}
\label{sec:conclusion}

In this paper we have given a detailed analysis of the semiclassical dynamics and spectrum of a spin 1 Bose microcondensate in the single mode approximation. This simple system proves to be rather rich, displaying Hamiltonian monodromy in the $\tilde q<0$ case and a separatrix that divides the phase space for $\tilde q>0$. Both of these classical phenomena have distinctive quantum analogues. 

It remains to make a few comments about the relation to recent experiments in ultracold gases. As mentioned in the introduction, several experiments have observed (semi-)classical dynamics consistent to some degree with the single mode approximation \cite{Chang:2005,black:2007,liu:2009} . Ref.~\cite{liu:2009} is of particular interest for its use of Faraday rotation spectroscopy in addition to the usual Stern--Gerlach separation of the different spin components. While the latter is only sensitive to the relative occupancies of the $m=+1,0,-1$ states, the former is capable of measuring the transverse magnetization, which in principal allows the rotation angle to be extracted (c.f. Fig.~\ref{fig:RotationLarge}). All of these experiments observe quite significant damping of single mode dynamics, indicating that other modes may be significant. That the system size be small compared to the spin healing length (equivalently, the level spacing in the trap is large compared to the spin interaction energy) is a necessary but probably insufficient criterion for the validity of the single mode approximation. The temperature is typically much larger than the level spacing, so that many modes are occupied. A reasonable expectation is that these `fast' modes adiabatically follow the slow dynamics of the condensate spinor, but a detailed theoretical description is presently lacking. Moving to still smaller systems could eliminate this complication.

Turning to the \emph{quantum} dynamics of a single mode, we note that the level spacing is set by the spin interaction energy $e_{2}$. Typical magnitudes are $4.3 \text{ Hz}$ for $^{87}\text{Rb}$ \cite{Chang:2005} and $33 \text{ Hz}$ for $^{23}\text{Na}$ \cite{liu:2009}. A direct observation of the quantized spectrum on these tiny energy scales seems unlikely at present. An alternative strategy is to ask how the dynamics is affected by this discreteness, leading to deviations from the classical predictions. The question was addressed in  several recent papers that treated related models \cite{babelon:2009,keeling:2009,faribault:2009}.


The author acknowledges the support of the NSF under grant DMR-0846788. 

\begin{appendix}
	
\section{Solution of the continuum Bethe equations}	\label{sec:bethecont}
	
The continuum limit of the Bethe Eqs.~\eqref{SpinorCondensateMonodromy_BAEqs} gives the integral equation for the root density $\rho(\lambda)$
\begin{multline}
	\label{SpinorCondensateMonodromy_bethecont}
	\text{P}\int  \frac{\rho(\lambda')}{\lambda'-\lambda}d\lambda'+1-\frac{\nu_{B}}{\lambda}-\frac{\nu_{K}}{\lambda+N\tilde q/2}=0\\\lambda\in \left\{\lambda:\rho(\lambda)\neq 0\right\}
\end{multline}
%
Write the root density as the jump in an analytic function
\[
	\rho(\lambda)=\frac{1}{2\pi i}\left[f(\lambda+i0)-f(\lambda-i0)\right],
\]
so that the integral in Eq.~\eqref{SpinorCondensateMonodromy_bethecont} can be viewed as circling the branch cuts of the $f(\lambda)$. If $f(\lambda)$ has the form
\begin{equation}
	\label{SpinorMonod_fform}
	f(\lambda)=\frac{\sqrt{P_{4}(\lambda)}}{\lambda(\lambda+N\tilde q/2)},
\end{equation}
in terms of a fourth order polynomial $P_{4}(\lambda)$ (required to get the two regions of nonzero root density) one can evaluate the integral in terms of the residues at $0$, $-N\tilde q/2$ and $\infty$, which solves the problem if 
\begin{equation}
	\label{P4conditions}
	\begin{split}
	\sqrt{P_{4}(0)}&=-\frac{\nu_{B}N\tilde q}{2}\\
	\sqrt{P_{4}(-N\tilde q/2)}&=\frac{\nu_{K}N\tilde q}{2}\\
	\sqrt{P_{4}(\lambda)}&\to\lambda^{2},\text{ as } \lambda\to\infty.
	\end{split}
\end{equation}
These conditions, together with a specification of the total number of roots and the energy
\begin{subequations}
	\label{SpinorMonod_NandXi}
\begin{gather}
	N_{R}=\int d\lambda\, \rho(\lambda)\label{Nquant}\\
	E = \frac{e_{2}}{2N}N(N-1)+q|S_{z}|-\frac{4e_{2}}{N}\int d\lambda\, \lambda\rho(\lambda),
\end{gather}
\end{subequations}
fix the form of $P_{4}(\lambda)$ uniquely. By evaluating Eq.~\eqref{SpinorMonod_NandXi} in the same way one can show that the polynomial arising from Eq.~\eqref{SpinorCondensateMonodromy_tevalue} solves the problem (in the large $N$ limit \texttt{The only thing that doesn't come out is the $\frac{1}{2}N_{R}(N_{R}-1)$} in Eq.~\eqref{SpinorMonod_BetheEnergy}).
A more useful approach to generating the spectrum is to separately quantize the number of roots in each of the two regions where the root density is nonvanishing, and then evaluating the resulting energy. The two quantization conditions are then identical to the WKB conditions Eq.~\eqref{SpinorMonod_WKBact}. The strength of the approach based on the Schr\"odinger equation is that the polynomial is explicitly given by Eq.~\eqref{SpinorCondensateMonodromy_tevalue} without the need to solve Eq.~\eqref{SpinorMonod_NandXi}. 

The observant reader may note that the form Eq.~\eqref{SpinorCondensateMonodromy_tevalue} is only consistent with the first two of the conditions Eq.~\eqref{P4conditions} in the limit of $\nu_{B,K}\gg 1$. The remedy is the Langer modification mentioned after Eq.~\eqref{SpinorMonod_WKBact}.
 	
\end{appendix}


%

\end{document}